\documentclass[usenatbib,useAMS]{mn2e}

\usepackage{psfig}

\newcommand{\Zsun}{\ensuremath{Z_{\odot}}}

\newcommand{\Hb}{\ensuremath{{\rm H}\beta}}
\newcommand{\CNone}{\ensuremath{{\rm CN}_1}}
\newcommand{\CNtwo}{\ensuremath{{\rm CN}_2}}
\newcommand{\FeC}{\ensuremath{{\rm C}_2}4668}
\newcommand{\Mgone}{\ensuremath{{\rm Mg}_1}}
\newcommand{\Mgtwo}{\ensuremath{{\rm Mg}_2}}
\newcommand{\Mgb}{\ensuremath{{\rm Mg}\, b}}
\newcommand{\TiOone}{\ensuremath{{\rm TiO}_1}}
\newcommand{\TiOtwo}{\ensuremath{{\rm TiO}_2}}
\newcommand{\Fe}{\ensuremath{\langle {\rm Fe}\rangle}}
\newcommand{\Teff}{\ensuremath{T_{\rm eff}}}
\newcommand{\aFe}{\ensuremath{\alpha/{\rm Fe}}}
\newcommand{\CaFe}{\ensuremath{{\rm Ca}/{\rm Fe}}}

\newcommand{\aCa}{\ensuremath{\alpha/{\rm Ca}}}

\newcommand{\aN}{\ensuremath{\alpha/{\rm N}}}
\newcommand{\FeH}{\ensuremath{{\rm Fe}/{\rm H}}}
\newcommand{\ZH}{\ensuremath{Z/{\rm H}}}
\newcommand{\BV}{\ensuremath{B\!-\!V}}
\newcommand{\VI}{\ensuremath{V\!-\!I}}

\title[Stellar population models with variable element abundance
ratios] {\boldmath Stellar population models of Lick indices with
variable element abundance ratios\thanks{Available in {\tt
people/dthomas/SSPs} at {\tt ftp.mpe.mpg.de}}}

\author[Daniel Thomas, Claudia Maraston, \& Ralf Bender] {Daniel
Thomas$^1$\thanks{Email: dthomas@mpe.mpg.de}, Claudia
Maraston$^1$, \& Ralf Bender$^{1,2}$\\ $^1$Max-Planck-Institut f\"ur
extraterrestrische Physik, Giessenbachstra\ss e, D-85748 Garching,
Germany\\ $^2$Universit\"ats-Sternwarte M\"unchen, Scheinerstr.~1,
D-81679 M\"unchen, Germany}

\date{Accepted 2002 November 4. Received {\ldots} ;
      in original form 2002 September 12}

\pagerange{\pageref{firstpage}--\pageref{lastpage}}

\pubyear{2002}

\begin{document}

\maketitle

\label{firstpage}

\begin{abstract}
We provide the whole set of Lick indices from CN$_1$ to TiO$_2$ in the
wavelength-range $4000\la\lambda\la 6500$~\AA\ of Simple Stellar
Population models with, for the first time, variable element abundance
ratios, $[\aFe]=0.0,\:0.3,\:0.5$, $[\aCa]=-0.1,\:0.0,\:0.2,\:0.5$, and
$[\aN]=-0.5,\:0.0$. The models cover ages between $1$ and $15$~Gyr,
metallicities between 1/200 and 3.5 solar. The impact from the element
abundance changes on the absorption-line indices are taken from
\citet{TB95}, using an extension of the method introduced by
\citet{Traetal00a}. Our models are free from the intrinsic \aFe\ bias
that was imposed by the Milky Way template stars up to now, hence they
reflect well-defined \aFe\ ratios at all metallicities.  The models
are calibrated with Milky Way globular clusters for which
metallicities and \aFe\ ratios are known from independent spectroscopy
of individual stars. The metallicities that we derive from the Lick
indices \Mgb\ and Fe5270 are in excellent agreement with the
metallicity scale by \citet{ZW84}, and we show that the latter
provides total metallicity rather than iron abundance.  We can
reproduce the relatively strong CN-absorption features \CNone\ and
\CNtwo\ of galactic globular clusters with models in which nitrogen is
enhanced by a factor three.  An enhancement of carbon, instead, would
lead to serious inconsistencies with the indices \Mgone\ and \FeC.
The calcium sensitive index Ca4227 of globular clusters is well
matched by our models with $[\CaFe]= 0.3$, including the metal-rich
Bulge clusters NGC~6528 and NGC~6553. From our \aFe\ enhanced models
we infer that the index [MgFe] defined by \citet{G93} is quite
independent of \aFe\ but still slightly decreases with increasing
\aFe. We find that the index ${\rm [MgFe]}^{\prime} \equiv\sqrt{\Mgb\
(0.72\cdot {\rm Fe5270}+0.28\cdot{\rm Fe5335})}$, instead, is
completely independent of \aFe\ and serves best as a tracer of total
metallicity.  Searching for blue indices that give similar information
as \Mgb\ and \Fe, we find that \CNone\ and Fe4383 may be best suited
to estimate \aFe\ ratios of objects at redshifts $z\sim 1$.
\end{abstract}

\begin{keywords}
stars: abundances -- Galaxy: abundances -- globular clusters: general
-- galaxies: stellar content -- galaxies: elliptical and lenticular,
cD

\end{keywords}


\section{Introduction}
\label{sec:intro}
The Lick system \citep{Buretal84,Fabetal85} defines absorption-line
indices at medium resolution ($\sim 8$~\AA) that can be used---through
the comparison with stellar population models---to derive ages and
metallicities of stellar systems.  Interestingly, the indices \Mgb\
and \Mgtwo\ of early-type galaxies yield higher metallicities (and
younger ages) than the indices Fe5270 and Fe5335 (\citealt{P89};
\citealt*{WFG92}; \citealt*{DSP93}; \citealt{CD94}; \citealt{BP95};
\citealt*{FFI95}; \citealt{Jor99}; \citealt{Mehetal98};
\citealt{Ku00}; \citealt{Lonetal00}; and others).  The most
straightforward qualitative interpretation of these strong Mg-indices
and/or weak Fe-indices is that the stellar populations in elliptical
galaxies have high Mg/Fe element ratios (or \aFe\ ratios if Mg is
taken as representative of $\alpha$-elements) with respect to the
solar values \citep{WFG92}. This finding strongly impacts on the
theory of galaxy formation, as super-solar \aFe\ ratios require short
star formation time-scales ($\la 1$~Gyr, \citealt{Ma94};
\citealt*{TGB99}), that are not achieved by current models of
hierarchical galaxy formation \citep{Th99a,TK99}.

However, there exist two major caveats about this conclusion. 1) Lick
indices have very broadly defined line windows ($\sim 40$~\AA). Each
index actually contains a large number of absorption features from
various elements, so that the direct translation into element
abundances is not very straightforward \citep*{Greggio97,TCB98}. 2)
The stellar library \citep{Woretal94} used in stellar population
models to compute Lick indices contains only very few stars with
metallicities above solar. An additional complication is that in these
libraries \aFe\ is not independent of \FeH\ (see
Section~\ref{sec:bias}).

In order to resolve these ambiguities, \citet{Maretal02} compare the
Lick indices of Simple Stellar Population (SSP) models with data of
metal-rich globular clusters of the Galactic Bulge \citep{Puzetal02},
the ages and and element abundances of which are known from
high-resolution stellar spectroscopy. They find that metal-rich, \aFe\
enhanced Bulge clusters show the same features as early-type galaxies:
their Mg indices are stronger than predicted by SSP models at a given
Fe index value.  This result is empirical evidence that Mg and Fe
indices indeed trace \aFe\ element ratios. \citet{Maretal02} verify
the uniqueness of interpreting the strong Mg indices and weak Fe
indices in elliptical galaxies in terms of Mg over Fe element
overabundance. They show that alternative explanations like
uncertainties in stellar evolution and SSP modelling, or a significant
steepening of the initial mass function (IMF) do either not reproduce
the observed indices or violate other observational constraints.
\citet{Maretal02} further show that the standard models reflect
variable element abundance ratios at the various metallicities, in
particular super-solar \aFe\ ratios at sub-solar metallicities.

Motivated by these results, we construct stellar population models for
various and well-defined element abundance ratios. We present the
whole set of Lick indices (\CNone, \CNtwo, Ca4227, G4300, Fe4383,
Ca4455, Fe4531, \FeC, \Hb, Fe5015, \Mgone, \Mgtwo, \Mgb, Fe5270,
Fe5335, Fe5406, Fe5709, Fe5782, Na~D, \TiOone, and \TiOtwo) of SSP
models with the \aFe\ ratios $[\aFe]=0.0,\:0.3,\:0.5$. The models
cover ages from $1$ to $15$~Gyr, and total metallicities from
$[\ZH]=-2.25$ to $0.65$. In these models, the elements nitrogen and
calcium are enhanced in lockstep with the other $\alpha$-elements,
hence $[\aN]=0.0$ and $[\aCa]=0.0$. Additionally, we provide models
with variable \aN\ and \aCa\ ratios, $[\aN]=-0.5$ and
$[\aCa]=-0.1,\:0.2,\:0.5$.  The impact from the element abundance
changes on the absorption-line indices are taken from \citet[][
hereafter TB95]{TB95}, using an extension of the method introduced by
\citet[][ hereafter T00]{Traetal00a}.  The models are calibrated with
the globular cluster data of \citet{Puzetal02}.  The present models
now allow for the unambiguous derivation of SSP ages, metallicities,
and element abundance, in particular \aFe, ratios.

The paper is organised as follows.  In Section~\ref{sec:construction}
we describe the construction of the models, and introduce the main
input parameters. In Section~\ref{sec:calibration} we present the
model results and their calibration with globular cluster data.  If
the reader is predominantly interested in the application of the
present SSP models, for the first reading we recommend to skip
Section~\ref{sec:construction}, and to focus on the summary given in
Section~\ref{sec:modelsum}.

\medskip
The models for selected ages are provided in the tables in the
appendix. Their complete versions are available electronically via ftp
at {\tt ftp.mpe.mpg.de} in the directory {\tt people/dthomas/SSPs},
via WWW at {\tt ftp://ftp.mpe.mpg.de/people/dthomas/SSPs}, or email to
{\tt dthomas@mpe.mpg.de}.


\section{Model Construction}
\label{sec:construction}
The classical input parameters for stellar population models are the
slope of the IMF, the age and the metallicity with fixed (solar)
element abundance proportions.  In this paper, we introduce the
abundances of individual elements as a further parameter, allowing for
various element mixtures at given total metallicity.  The new SSP
models are based on the standard SSP models computed with the code of
\citet{Ma98}. For Lick indices refer to \citet{MT00}, \citet{MGT01},
and \citet{Maretal02}. These base models are modified according to the
desired element abundance ratios. In the following paragraphs of this
section we describe these modifications step by step and introduce the
main input parameters.

\subsection{The basic SSP model}
The underlying SSP models are presented in \citet{Ma98} and
C.~Maraston (in preparation). In these models, the fuel consumption
theorem \citep{RB86} is adopted to evaluate the energetics of the post
main sequence phases. The input stellar tracks (solar abundance
ratios) with metallicities from 1/200 to 2 solar, are taken from
\citet*{CCC97}, \citet{Betal97}, and S.~Cassisi (1999, private
communication). The tracks with 3.5 solar metallicity are taken from
\citet{Saletal00}. Lick indices are computed with the calibrations of
the indices as functions of the stellar parameters (the so-called
fitting functions) by \citet{Woretal94}. The impact from using
alternatively the fitting functions of Buzzoni \citep*{BGM92,BMG94} or
\citet{Boetal95} and the resulting uncertainties in the modelling are
discussed in \citet*{MGT01} and \citet{Maretal02}. In this paper we
focus on the fitting function of \citet{Woretal94}, because they
comprise all 21 Lick absorption line indices. Our models adopt a
\citet{Salpeter55} IMF slope.

\subsection{Varying element abundances}
\label{sec:abundancevar}
The aim is to obtain SSP models with various and well-defined element
abundance ratios at fixed total metallicity.  The most important ratio
is \aFe, which is the ratio of the so-called $\alpha$-elements (N, O,
Mg, Ca, Na, Ne, S, Si, Ti) to the Fe-peak elements (Cr, Mn, Fe, Co,
Ni, Cu, Zn), because it carries information on the formation time-scale
of stellar populations (see the Introduction).

\subsubsection{Enhanced and depressed groups}
To enhance the abundance ratio \aFe\ keeping the total metallicity
constant, the increase in the abundances of the $\alpha$-elements has
to be counterbalanced by the decrease of Fe-peak element
abundances. Following T00's notation, we call the former enhanced
group and the latter depressed group. We keep the abundance of carbon
fixed, as in the solar neighbourhood carbon appears less enhanced than
the $\alpha$-elements \citep{McW97}. T00 include a model in which
carbon is assigned to the depressed group. We tested this option, and
found that the depression of carbon leads to serious inconsistencies
between models and globular cluster data of the indices \CNone,
\CNtwo, \FeC, and \Mgone\ (see Section~\ref{sec:calibration}, T00).
All elements heavier than Zn are assumed not to vary.

Our prescriptions are identical to T00's Model~1, except that we
include the $\alpha$-element calcium in the enhanced group.  This
choice is motivated by the evidence that in halo and disc of our
Galaxy the element calcium follows the typical abundance patterns of
the other $\alpha$-elements like oxygen and magnesium \citep{McW97}.
T00, instead, assign calcium to the depressed elements, because
elliptical galaxies have low Ca4227 and Ca4455 indices
\citep{Vazetal97,Wor98,Traetal98}.  In a separate model we explored
this case.  We verified that the resulting SSP models of none of the
21 indices---except Ca4227---are significantly different when calcium
is assigned either to the enhanced or the depressed group, simply
because only Ca4227 is sensitive to calcium abundance
(TB95). Curiously, Ca4455 is completely insensitive to the element
calcium (TB95). Moreover, the fractional contribution of Ca to the
total metallicity is too small ($\sim 0.1$ per cent) to change the
isochrone and SSP characteristics.

\subsubsection{Varying \aFe\ ratios at constant metallicity}

The starting point is a total metallicity in which the mass fractions
of the enhanced and depressed groups are $X^+$ and $X^-$,
respectively. Its \aFe\ ratio normalized to the solar value
$(X^+_{\odot}/X^-_{\odot})$ is defined as
\begin{equation}
[\aFe]=\log\left(\frac{X^+}{X^-}\right) -
\log\left(\frac{X^+_{\odot}}{X^-_{\odot}}\right)\ .
\label{eqn:1}
\end{equation}
In order to obtain a new chemical mixture with the same total
metallicity, we change $X^+$ and $X^-$ by the factors $f_{\alpha}$ and
$f_{\rm Fe}$, respectively, such that the new \aFe\ is
\begin{equation}
[\aFe]_{\rm new} = \log\left(\frac{f_{\alpha}}{f_{\rm Fe}}\right) +
\log\left(\frac{X^+}{X^-}\right) -
\log\left(\frac{X^+_{\odot}}{X^-_{\odot}}\right)\ .
\label{eqn:2}
\end{equation}

The conservation of total metallicity requires the following condition
\[
f_{\alpha}\ X^{+} + f_{\rm Fe}\ X^{-} = X^{+} + X^{-}\ ,
\]
or, dividing by $X^-$,
\begin{equation}
f_{\alpha} \left(\frac{X^{+}}{X^-}\right) + f_{\rm Fe} = \left(\frac{X^{+}}{X^-}\right) + 1\ .
\label{eqn:3}
\end{equation}
From Equations~\ref{eqn:1} to \ref{eqn:3} the factors $f_{\alpha}$ and
$f_{\rm Fe}$ can be determined for given [\aFe] and $[\aFe]_{\rm
new}$.

From \citet*{GNS96} we adopt: $X^{+}_{\odot}=0.0148$ (solar oxygen
abundance $X^{\rm O}_{\odot}=0.0096$), $X^{-}_{\odot}=0.0016$, and
total solar metallicity $\Zsun=0.02$. With these values, a new
abundance ratio $[\aFe]_{\rm new}=0.3$ starting from the solar values
$[\aFe]=0$ (i.e. increasing \aFe\ by a factor of 2) is obtained with
$f_{\alpha}=1.052$ and $f_{\rm Fe}=0.526$. Thus, as also emphasized by
T00, super-solar \aFe\ ratios at fixed total metallicity are produced
by a decrease of the Fe abundance rather than by an increase of the
$\alpha$-element abundances. The reason for this effect is that the
$\alpha$-element oxygen is the most abundant element in the Sun after
hydrogen and helium, so that total metallicity is by far dominated by
oxygen (48 per cent by mass). The remaining $\alpha$-elements
contribute 26 per cent, the Fe-peak elements only 8 per cent to the
total amount of metals in the Sun.

The following notation will be used throughout this paper.  We
distinguish between total metallicity [\ZH], iron abundance [\FeH],
and $\alpha$-element to iron ratio [\aFe]. Only two of these
quantities are independent. Following \citet{TCB98} and T00, they can
be related through the following equation.
\begin{equation}
[\ZH]=[\FeH]+A\ [\aFe]
\label{eqn:Fescaling}
\end{equation}
with 
\[
A=-\frac{\Delta[\FeH]}{\Delta[\aFe]}
\]
The factor $A$ depends on the partition between enhanced and depressed
elements.  For our adopted mixtures (see previous section) we obtain
$A=0.94$. For more details see T00.

\subsubsection{Varying \aN\ and \aCa\ ratios}
Both nitrogen and calcium are assigned to the enhanced
group. Therefore, the ratios \aN\ and \aCa\ are fixed to the solar
value ($[\aN]=0,\ [\aCa]=0$) in the \aFe\ enhanced mixtures described
in the previous section. We computed additional models with chemical
mixtures in which nitrogen and calcium are detached from the group of
enhanced elements and their abundances are allowed to vary with
respect to the abundances of the other $\alpha$-elements. We computed
a model in which nitrogen is enhanced by a factor 3 with respect to
the other $\alpha$-elements, hence $[\aN]=-0.5$, and various models in
which calcium is depressed with respect to the other
$\alpha$-elements, hence $[\aCa]=0.1, -0.2, -0.5$. Note that the
fractional contribution of the elements nitrogen and calcium to total
metallicity is only $\sim 5$ and 0.3 per cent, respectively. The
different \aN\ and \aCa\ ratios are therefore achieved essentially by
changing the abundances of the elements N and Ca, and the perturbation
of the total metallicity budget is negligible.

\subsection{Effects of element abundances on Lick indices}
\label{sec:responsemodel}
The account for the impact from abundance variations of individual
elements on the absorption line-strengths is the principal ingredient
of models with variable element abundances. In the models of this
paper, the variation of the Lick absorption line indices owing to
element abundance changes is taken from TB95 as described below.

\subsubsection{Effect on representative evolutionary phases}
\label{sec:phases}
\begin{figure}
\psfig{file=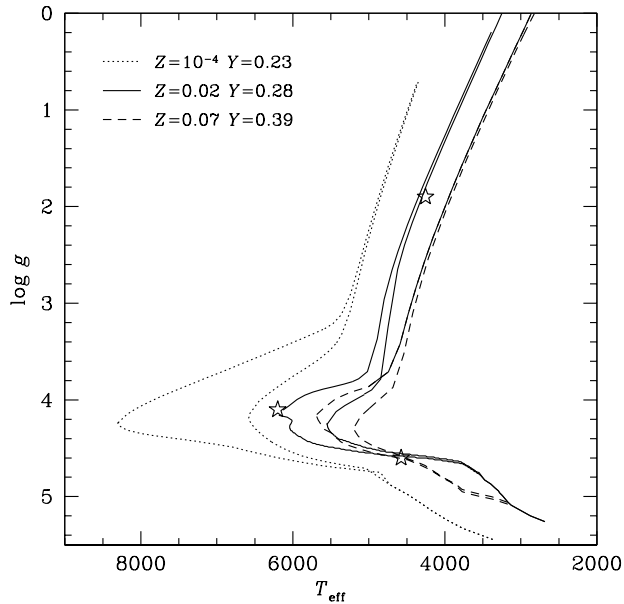,width=\linewidth}
\caption{Dependence of the position of the reference model by
\citet{TB95} on metallicity and age. The star symbols are the
representative positions of the three evolutionary phases dwarfs,
turnoff and giants (from bottom to top) used in
\citet{TB95}. Isochrones with solar (solid line) and sub-solar (dotted
lines) metallicity are from S.\ Cassisi (see text), the very
metal-rich ones (dashed lines) are taken from Salasnich et
al.~(2000). For every metallicity, models with 5 and 15 Gyr are
plotted, to show also the effect of the age. A temperature of 5000~K
is adopted to separate cool dwarfs from turnoff stars on the main
sequence, at every metallicity.}
\label{fig:tbphases}
\end{figure}

TB95 computed model atmospheres and synthetic spectra along a 5~Gyr
old isochrone with solar metallicity.  On these model atmospheres they
assess the impact on Lick indices from element abundance variations.
The model atmospheres have well-defined values of temperature and
gravity, chosen to be representative of the three evolutionary phases,
dwarfs ($\Teff=4575~K$, $\log g=4.6$), turnoff ($\Teff=6200~K$, $\log
g=4.1$) and giants ($\Teff=4255~K$, $\log g=1.9$). These couples of
\Teff\ and $\log g$ are very appropriately chosen. The giant model,
for instance, is placed at a location of the isochrone where most of
the fuel is burned (Fig.~\ref{fig:tbphases}), i.e.\ at the base of the
Red Giant Branch and on the Horizontal Branch \citep{Ma98}. For
instance, a reference model for giants with a much lower gravity would
have overestimated the effect on the Mg indices, which are very strong
at very low gravity and temperature.

At different metallicities and ages, the three evolutionary phases are
in principle represented by slightly different couples of \Teff\ and
$\log g$. The effect is shown in Fig.~\ref{fig:tbphases}, where
couples of isochrones (with ages 5 and 15 Gyr) at various
metallicities (see references in the caption) are plotted in the
\Teff\ versus $\log g$ plane.  The star symbols denote the dwarfs,
turnoff and giants locations defined by TB95. The larger the
metallicity, the closer is the dwarf-border to the turnoff, and vice
versa. The turnoff is hotter at decreasing age or metallicity, but the
gravity keeps rather constant around $\log g\sim
4.1$. Fig.~\ref{fig:tbphases} shows that the phases are rather well
defined independent of age and metallicity. We use the fixed
temperature of 5000~K to separate the turnoff region from the cool
dwarfs on the Main Sequence, independent of age and metallicity. In
order to assess quantitatively the impact of this choice, we computed
the integrated indices of the most metal-rich SSP adopting 4000~K
instead of 5000~K.  We find that the resulting SSP indices change by
only $\sim 0.1$ per cent. The impact is so small because for standard
IMFs dwarfs with these temperatures play a minor role in the
integrated indices \citep{Maretal02}.  We assign the Subgiant Branch
phase to the turnoff because of the very similar \Teff\ and $\log
g$. The evolutionary phase `giants' consists of the Red Giant Branch,
the Horizontal Branch, and the Asymptotic Giant Branch phases.

\subsubsection{Tripicco \& Bell Response Functions}
On each model atmosphere---dwarfs, turnoff, giants---TB95 measure the
absolute Lick index value $I_0$. Doubling in turn the abundances $X_i$
of the dominant elements C, N, O, Mg, Fe, Ca, Na, Si, Cr, and Ti, they
determine the index changes $\Delta I(i)$.  The abundance effects are
therefore isolated at a given temperature and surface gravity.  In
this way, TB95 provide the first partial derivative $\partial
I/\partial [X_i]$ of the index $I_0$ for the logarithmic element
abundance increment $\Delta [X_i]\equiv\log X_i^1/X_i^0=\log 2=0.3\
{\rm dex}$.  Information on the second partial derivatives is not
provided by the TB95 calculations.

The most simple approach to express the index $I$ as a function of
element abundances is to use a Taylor series:
\begin{equation}
I_{\rm new} = I + \sum_{i=1}^{n} \frac{\partial I}{\partial [X_i]}\; \Delta [X_i]
+\ \ldots\ ,
\label{eqn:lineartaylor}
\end{equation}
with $i$ indicating a chemical element.

The logarithmic abundance variation $\Delta [X_i]$ depends on the
desired element abundance ratio, and is determined for the elements of
the enhanced and the depressed groups separately from
Eqns.~\ref{eqn:2} and \ref{eqn:3}. Hence $\Delta [X_{i}]=\log
f_{\alpha}$ and $\Delta [X_{i}]=\log f_{\rm Fe}$ for elements from the
enhanced and depressed group, respectively.
Equation~\ref{eqn:lineartaylor} can be applied, only if the second
(and higher order) derivatives are negligible. In other words, the
absorption line indices must be linear functions of the logarithm of
the element abundances, hence $I\propto [X_i]$. Is this the case?

Studies on the behaviour of the Lick indices as functions of
individual element abundances are only available from the work by
TB95.  However, it is reasonable to assume that in the linear regime
of the growth curve the absorption line strength $I$ is proportional
to the number of absorbers, so that we can expect $I\propto X_i\propto
\exp ([X_i])$. Hence, Equation~\ref{eqn:lineartaylor}, that requires
instead $I\propto [X_i]$, is not an optimal approximation. This rough
estimate gets support from the work by \citet{Boetal95}, who derive
empirically from Milky Way stars that $\Mgtwo\propto\exp([{\rm
Mg}/{\rm H}])$ (the other Lick indices are not discussed in
\citealt{Boetal95}). This functional form implies that---at least in
the linear part of the growth curve---$\ln I$ instead of $I$ is a
linear function of $[X_i]$. Therefore, it is more appropriate to
consider the Taylor expansion of $\ln I$ in place of
Eqn.~\ref{eqn:lineartaylor}.
\begin{equation}
\ln I_{\rm new} = \ln I + \sum_{i=1}^{n} \frac{\partial \ln I}{\partial
[X_i]}\; \Delta [X_i] +\ \ldots
\label{eqn:lntaylor}
\end{equation}
Neglecting the higher order derivatives we can write
\begin{eqnarray}
\ln I_{\rm new} &=& \ln I + \sum_{i=1}^{n} \frac{\partial \ln I}{\partial
[X_i]}\; \Delta [X_i]\nonumber\\ 
&=& \ln I + \sum_{i=1}^{n} \frac{1}{I_0}\frac{\partial I}{\partial
[X_i]}\; 0.3\; \frac{\Delta [X_i]}{0.3}\nonumber\\
&=& \ln I + \sum_{i=1}^{n}\ R_{0.3}(i)\; \frac{\Delta [X_i]}{0.3}\nonumber\ ,
\end{eqnarray}
with $R_{0.3}(i)$---following the notation of T00---being the index
response to the abundance change of the element $i$ by 0.3 dex
(hereafter, specific fractional index change) given in TB95. Taking the
exponential we obtain
\begin{eqnarray}
I_{\rm new} &=& I\ \prod_{i=1}^{n}\:\exp
\left\{\frac{1}{I_0}\frac{\partial I}{\partial
[X_i]}\; 0.3\right\}^{\left(\Delta [X_i]/0.3\right)}\nonumber\\
 &=& I\ \prod_{i=1}^{n}\:\exp
\left\{R_{0.3}(i)\right\}^{\left(\Delta [X_i]/0.3\right)}
\label{eqn:exp}
\end{eqnarray}
A further approximation of the exponential $\exp
\left\{R_{0.3}(i)\right\}$ to $1+R_{0.3}(i)$ assuming that
$R_{0.3}(i)\ll 1$ yields the formula introduced in T00. As the
specific fractional index change can be as large as $R_{0.3}(i)\approx
0.2-0.4$, we will not adopt this approximation, but use instead
Eqn.~\ref{eqn:exp} for the computation of our models.
The elements included are those considered in TB95, i.e.\ C, N, O, Mg,
Ca, Na, Si, Ti, Fe, and Cr.

\subsubsection{Negative indices}
\label{sec:negative}
The approach of expanding $\ln I$ (Equation~\ref{eqn:lntaylor})
assumes
\[ 
I= \textrm{const.}\cdot e^{\: [X_i]}\ ,
\]
which implies that the index approaches asymptotically the value zero
for very low element abundances, i.e.\ $I\rightarrow 0$ for
$X_i\rightarrow 0$ or $[X_i]\rightarrow -\infty$ (see also T00). This
condition, however, is not generally fulfilled for Lick indices. They
can become negative, depending on the definition of line and
pseudo-continuum windows. This typically happens at young ages and/or
low abundances, and must be corrected before using Eqn.~\ref{eqn:exp},
otherwise $\ln I$ is not defined.  We therefore shift negative index
values in Eqn.~\ref{eqn:exp} by the amount $\delta$ required to
approach zero at zero element abundances.  To mimic the (unknown)
index value at zero element abundance, we set
\begin{equation}
\delta\equiv I_{\rm low}-|I_{\rm low}|\ .
\label{eqn:delta}
\end{equation}
$I_{\rm low}$ is the index value at the lowest metallicity of our grid
at a given age, i.e.\ $I_{\rm low}=I([\ZH]=-2.25)$ for metallic
indices.  For \Hb\ we use $I_{\rm low}=I([\ZH]=0.67)$.

Then we apply the TB95 correction to the scaled index $I-\delta$.
After the TB95 correction we scale the resulting index $I_{\rm
new}-\delta$ back. This procedure can be summarized in the following
modification of Equation~\ref{eqn:exp}:
\begin{equation}
I_{\rm new}-\delta = (I-\delta)\ \prod_{i=1}^{n}\:\exp
\left\{\frac{1}{I_0-\delta}\frac{\partial I}{\partial [X_i]}\;
0.3\right\}^{\left(\Delta [X_i]/0.3\right)}
\label{eqn:modexp}
\end{equation}

Indices with positive values are assumed to reach the value zero at
zero abundances and therefore do not require any correction, i.e.\
$\delta=0$.

We determine $\delta$ for each evolutionary phase of each SSP
separately. These $\delta$ values for the illustrative case of a
12~Gyr isochrone are shown in Table~\ref{tab:shift}. Only the indices
\CNone, \CNtwo, \FeC, and Fe5782 are significantly affected.  The
classical indices \Mgtwo, \Mgb, Fe5270, Fe5335, and \Hb\ (the
contribution from the dwarf phase to \Hb\ is negligible), do not
require any correction.
\begin{table}
\begin{minipage}{0.6\linewidth}
\caption{$\delta$ corrections to negative indices for a 12 Gyr
isochrone}
\begin{tabular}{lccc}
\hline\hline
\multicolumn{1}{c}{Index} & \multicolumn{1}{c}{Dwarf} &
\multicolumn{1}{c}{Turnoff} & \multicolumn{1}{c}{Giant}\\
\hline
\CNone   & $-0.344$ & $-0.309$ & $-0.119$\\
\CNtwo   & $-0.296$ & $-0.204$ & $-0.057$\\
Ca4227   & -- & -- & --\\
G4300    & -- & $-1.405$ & --\\
Fe4383   & -- & $-1.116$ & --\\
Ca4455   & -- & -- & --\\
Fe4531   & -- & -- & --\\
\FeC     & $-5.945$ & $-1.344$ & $-0.024$\\
\Hb      & $-0.974$ & -- & --\\
Fe5015   & -- & -- & --\\
\Mgone   & -- & $-0.019$ & $-0.009$\\
\Mgtwo   & -- & -- & --\\
\Mgb     & -- & -- & --\\
Fe5270   & -- & -- & --\\
Fe5335   & -- & -- & --\\
Fe5406   & -- & -- & --\\
Fe5709   & $-0.030$ & -- & --\\
Fe5782   & $-0.468$ & $-0.316$ & $-0.267$\\
Na~D     & -- & -- & --\\
\TiOone  & -- & -- & --\\
\TiOtwo  & $-0.028$ & $-0.004$ & $-0.004$\\
\hline
\end{tabular}
\label{tab:shift}
\end{minipage}
\end{table}

\begin{table*}
\begin{minipage}{0.88\textwidth}
\caption{Index Values and Fractional Index Changes for $t=12$~Gyr and $[\aFe]=0.3$}
\begin{tabular}{lrrrrrrrrrrrr}
\hline\hline 
& \quad & \multicolumn{3}{c}{$I_0$ TB95} &\quad &
 \multicolumn{3}{c}{$I_0$ This Work}& \quad & \multicolumn{3}{c}{$\Delta I/(I_0-\delta)$}\\
 \multicolumn{1}{c}{Index} & & \multicolumn{1}{c}{Dwarf} &
 \multicolumn{1}{c}{Turnoff} & \multicolumn{1}{c}{Giant} & &
 \multicolumn{1}{c}{Dwarf} & \multicolumn{1}{c}{Turnoff} &
 \multicolumn{1}{c}{Giant} & & \multicolumn{1}{c}{Dwarf} &
 \multicolumn{1}{c}{Turnoff} & \multicolumn{1}{c}{Giant}\\
 \multicolumn{1}{c}{(1)} & &
 \multicolumn{1}{c}{(2)} & \multicolumn{1}{c}{(3)} &
 \multicolumn{1}{c}{(4)} & & \multicolumn{1}{c}{(5)} &
 \multicolumn{1}{c}{(6)} & \multicolumn{1}{c}{(7)} & &
 \multicolumn{1}{c}{(8)} & \multicolumn{1}{c}{(9)} &
 \multicolumn{1}{c}{(10)}\\
\hline
\CNone   & & $ 0.04$ & $-0.09$ & $ 0.28$ & & $ 0.00$ & $-0.07$ & $ 0.18$ & & $ 0.086$ & $-0.007$ & $ 0.079$\\
\CNtwo   & & $ 0.12$ & $-0.06$ & $ 0.39$ & & $ 0.04$ & $-0.04$ & $ 0.23$ & & $ 0.085$ & $-0.012$ & $ 0.096$\\
Ca4227   & & $ 5.33$ & $ 0.58$ & $ 3.39$ & & $ 3.09$ & $ 0.80$ & $ 1.40$ & & $ 0.058$ & $-0.078$ & $ 0.000$\\
G4300    & & $ 6.83$ & $ 2.98$ & $ 8.00$ & & $ 5.37$ & $ 3.96$ & $ 6.51$ & & $ 0.091$ & $ 0.030$ & $ 0.068$\\
Fe4383   & & $10.09$ & $ 1.61$ & $ 9.02$ & & $ 6.98$ & $ 3.07$ & $ 7.01$ & & $-0.315$ & $-0.113$ & $-0.207$\\
Ca4455   & & $ 1.89$ & $ 0.47$ & $ 2.28$ & & $ 2.38$ & $ 1.02$ & $ 2.10$ & & $-0.011$ & $-0.045$ & $ 0.035$\\
Fe4531   & & $ 4.28$ & $ 1.76$ & $ 3.59$ & & $ 4.52$ & $ 2.39$ & $ 4.06$ & & $-0.096$ & $-0.074$ & $-0.050$\\
\FeC     & & $ 2.34$ & $ 0.77$ & $ 8.62$ & & $ 4.19$ & $ 2.35$ & $ 7.48$ & & $-0.033$ & $-0.050$ & $-0.031$\\
\Hb      & & $-0.10$ & $ 3.79$ & $ 0.05$ & & $ 0.09$ & $ 2.91$ & $ 1.34$ & & $ 0.030$ & $ 0.033$ & $ 0.006$\\
Fe5015   & & $ 2.72$ & $ 2.42$ & $ 4.79$ & & $ 5.11$ & $ 3.77$ & $ 6.64$ & & $-0.071$ & $-0.015$ & $-0.063$\\
\Mgone   & & $ 0.33$ & $ 0.01$ & $ 0.25$ & & $ 0.24$ & $ 0.02$ & $ 0.12$ & & $ 0.176$ & $ 0.126$ & $ 0.204$\\
\Mgtwo   & & $ 0.53$ & $ 0.07$ & $ 0.36$ & & $ 0.44$ & $ 0.11$ & $ 0.27$ & & $ 0.076$ & $ 0.036$ & $ 0.093$\\
\Mgb     & & $ 7.12$ & $ 1.25$ & $ 3.65$ & & $ 5.40$ & $ 2.30$ & $ 3.81$ & & $ 0.170$ & $ 0.078$ & $ 0.231$\\
Fe5270   & & $ 4.79$ & $ 1.31$ & $ 4.49$ & & $ 3.83$ & $ 1.84$ & $ 3.50$ & & $-0.170$ & $-0.096$ & $-0.130$\\
Fe5335   & & $ 4.05$ & $ 0.93$ & $ 3.40$ & & $ 3.56$ & $ 1.57$ & $ 3.09$ & & $-0.243$ & $-0.152$ & $-0.206$\\
Fe5406   & & $ 3.01$ & $ 0.63$ & $ 2.60$ & & $ 2.30$ & $ 0.82$ & $ 2.15$ & & $-0.254$ & $-0.278$ & $-0.192$\\
Fe5709   & & $ 1.01$ & $ 0.37$ & $ 1.59$ & & $ 0.78$ & $ 0.69$ & $ 1.19$ & & $-0.173$ & $-0.155$ & $-0.092$\\
Fe5782   & & $ 0.68$ & $ 0.16$ & $ 1.12$ & & $ 0.83$ & $ 0.41$ & $ 1.07$ & & $-0.095$ & $-0.133$ & $-0.131$\\
Na~D     & & $ 8.11$ & $ 0.66$ & $ 3.93$ & & $ 7.19$ & $ 1.52$ & $ 3.31$ & & $ 0.035$ & $-0.092$ & $ 0.033$\\
\TiOone  & & $ 0.00$ & $-0.01$ & $ 0.01$ & & $ 0.07$ & $ 0.01$ & $ 0.05$ & & $ 0.002$ & $-0.104$ & $ 0.046$\\
\TiOtwo  & & $ 0.02$ & $ 0.01$ & $ 0.05$ & & $ 0.27$ & $ 0.00$ & $ 0.09$ & & $-0.010$ & $-0.483$ & $-0.024$\\
\hline
\end{tabular}
\label{tab:indices}
\end{minipage}
\end{table*}

\subsubsection{Total fractional index changes}
\label{sec:fract}
The specific (i.e.\ referred to element $i$) fractional index change
\[
R_{0.3}(i)=\frac{1}{I_0}\frac{\partial I}{\partial [X_i]}\; 0.3\ ,
\]
is the main input in Equation~\ref{eqn:exp}. TB95 provide both $I_0$
and $(\partial I/\partial [X_i])\cdot 0.3$. However, the authors do
not always match well the $I_0$ values measured for Milky Way
stars. One of the striking examples is \Hb\ in cool dwarfs and
giants. Owing to the neglect of non-LTE effects (TB95), TB95 measure
on their model atmosphere with $T_{\rm eff}=4255$~K and $\log g=1.9$
the absorption index $\Hb=0.05$~\AA, while cool giants with that
temperature typically have $\Hb>1$~\AA\ (see Fig.~12 in TB95). This
discrepancy results in a higher fractional response $R_{\rm 0.3}$ by a
factor of 20. For the dwarfs TB95 obtain $\Hb=-0.1$~\AA, while dwarfs
with $T_{\rm eff}= 4600$~K have $\Hb>0$ (Fig.~12 in TB95).  We
therefore prefer to rely on the values provided by TB95 in a
differential sense, and adopt from TB95 only the index variations
($(\partial I/\partial [X_i])\times 0.3$ in Eqn.~\ref{eqn:exp}).  The
absolute $I_0$ values for the three evolutionary phases, instead, are
those of our underlying 5~Gyr, \Zsun\ SSP model. These values and the
original $I_0$ values of TB95 are listed in Columns~$2-7$ of
Table~\ref{tab:indices}.  It can be seen that the difference between
our and TB95's $I_0$ values is significant for the indices \CNone,
\CNtwo, \TiOone, and \TiOtwo\ in the dwarf phase, Ca4455, \FeC,
Fe5782, Na~D, \TiOone, and \TiOtwo\ in the turnoff phase, and Ca4227,
\Hb\ (see above), \Mgone, and \TiOone\ in the (dominating) giant
phase. Again, the frequently used indices \Mgtwo, \Mgb, Fe5270, and
Fe5335 are very well matched by the models of TB95.

The total (i.e.\ integrated over all elements) fractional index
changes in each phase obtained from applying Eqn.~\ref{eqn:modexp} are
given in the Columns~8--10 of Table~\ref{tab:indices}. These numbers
give the percentage variations of all individual indices to an
increase of the \aFe\ ratio to $[\aFe]=0.3$. They are the key
ingredient in our \aFe\ enhanced models. The 21 indices can be roughly
divided into three groups. Those showing significant positive
responses to \aFe\ enhancement are: \CNone, \CNtwo, \Mgone, \Mgtwo,
and \Mgb. Significant negative responses, instead, are displayed by
Fe4383, Fe4531, \FeC, Fe5015, Fe5270, Fe5335, Fe5406, Fe5709, and
Fe5782. The indices Ca4227, G4300, Ca4455, \Hb, Na~D, \TiOone, and
\TiOtwo, instead, appear almost insensitive to the \aFe\ element
abundance ratio changes. The indices with the strongest fractional
responses of the order $\sim 20$~per cent are Fe4383, \Mgone, \Mgb,
Fe5335, and Fe5406.

\subsubsection{The final step}
The final model index $I_{\rm new}^{\rm SSP}$ is now computed in the
following way.  The basic SSP model is split in the three evolutionary
phases, dwarfs (D), turnoff (T) and giants (G), as explained above. We
compute the Lick indices of the base model for each phase separately,
and modify them according to Eqns.~\ref{eqn:exp} or~\ref{eqn:modexp}
using the fractional responses (like given in Table~\ref{tab:indices}
for $t=12$~Gyr).  The new total integrated index of the SSP is then
\begin{equation} 
I_{\rm new}^{\rm SSP} = \frac{I_{\rm new}^{\rm D} \times F_{\rm
C}^{\rm D} + I_{\rm new}^{\rm T} \times F_{\rm C}^{\rm T} + I_{\rm
new}^{\rm G} \times F_{\rm C}^{\rm G}}{F_{\rm C}^{\rm D} + F_{\rm
C}^{\rm T} + F_{\rm C}^{\rm G}}\ ,
\label{eqn:sspus}
\end{equation}
where $I_{\rm new}^{\rm D}$, $I_{\rm new}^{\rm T}$, $I_{\rm new}^{\rm
G}$ are the integrated indices in the three phases after TB95
correction, and $F_{\rm C}^{\rm D}$, $F_{\rm C}^{\rm T}$, $F_{\rm
C}^{\rm G}$ are their continua. It can be easily verified that
Eqn.~\ref{eqn:sspus} is mathematically equivalent to
\begin{equation}
I_{\rm SSP} = \Delta \left(1-\frac{\sum_i F^i_{\rm L}}{\sum_i F^i_{\rm
C}}\right)\ ,
\label{eqn:ssp}
\end{equation}
which defines integrated indices (in EW) of SSPs. In
Eqn.~\ref{eqn:ssp} $F^i_{\rm L}$ and $F^i_{\rm C}$ are the fluxes in
the line and the continuum (of the considered index), for the $i$-th
subphase of the population, $\Delta$ is the line width
\citep[see][]{Maretal02}.

TB95 performed their exercise only for a 5~Gyr, \Zsun\ isochrone, so
that we have to assume that the fractional responses are independent
of age and metallicity. We expect only small age dependencies, because
the spectrophotometric evolution of an SSP is mild after the RGB phase
transition ($\sim 1-2$~Gyr), hence for the ages computed for the
models of this paper.  The dependency on metallicity needs to be
assessed through TB95-calculations at low element abundances. Such a
detailed assessment, which goes by far beyond the scope of this paper,
is subject of future investigations.

\subsection{\boldmath Correcting the $\alpha$/Fe bias of stellar libraries}
\label{sec:bias}
In stellar population models, the link between Lick absorption line
indices and the stellar parameters temperature, gravity, and
metallicity is provided by the so-called fitting functions. These are
constructed from empirical stellar libraries that reflect the chemical
history of the Milky Way. Metal-poor halo stars ($[\FeH]\la -1$) have
$[\aFe]\approx 0.3$, because they formed at early epochs when the
chemical enrichment was dominated by Type II supernovae
nucleosynthesis. The \aFe\ ratios of the metal-rich disk stars,
instead, decreases from $[\aFe]\approx 0.3$ to solar for increasing
metallicity $-1\la [\FeH]\la 0$ \citep{EAGLNT93,Fu98} owing to the
delayed Type Ia supernova enrichment
\citep*[e.g.,][]{GR83,MG86,PT95,TGB98}.  This implies that every
stellar population model adopting these Milky Way based index
calibrations suffers from this bias in the \aFe\ ratio, i.e.\ the
model Lick indices reflect super-solar \aFe\ at sub-solar
metallicities \citep{Boetal95}. This is confirmed by the calibration
of the standard models in \citet{Maretal02}.

\begin{table}
\caption{The \aFe\ Bias in the Milky Way}
\begin{center}
\begin{tabular}{lrrrrrr}
\hline\hline
 {[\ZH]} & $-2.25$ & $-1.35$ & $-0.33$ & $0.00$ & $0.35$ & $0.67$\\
 {[\aFe]} & $0.25$ &  $0.20$ &  $0.10$ & $0.00$ & $0.00$ & $0.00$\\
\hline
\end{tabular}
\end{center}
\label{tab:bias}
\end{table}
Therefore, we assume that the base model does not have solar abundance
ratios at every metallicity, but possess this residual intrinsic \aFe\
ratio, or \aFe\ bias, at sub-solar metallicities. In our models we
account for this bias with the aid of Eqn.~\ref{eqn:1} to \ref{eqn:3},
putting as the starting [\aFe] in Eqn.~\ref{eqn:1} the \aFe\ bias. The
adopted values of [\aFe] for the various metallicities are given in
Table~\ref{tab:bias}. These are based on abundance measurements in
Milky Way stars (see the review by \citealt{McW97} and references
therein) and are those providing the best calibration of our resulting
SSP models with globular cluster (GC) data (see
Section~\ref{sec:calibration}).  We further assume that the input
[\FeH] in the fitting functions is not total metallicity but iron
abundance \citep[see][]{Maretal02}.  Thanks to these corrections we
are able to provide for the first time SSP models with well-defined
\aFe\ ratios at all metallicities.

\subsection{Summary of model construction}
\label{sec:modelsum}
\paragraph*{Base model.---}The 
underlying SSP models are presented in \citet{Ma98} and
\citet{Maretal02}. In these models, the fuel consumption theorem
\citep{RB86} is adopted to evaluate the energetics of the post main
sequence phases. The input stellar tracks (solar abundance ratios)
with metallicities from 1/200 to 2 solar, are taken from
\citet{CCC97}, \citet{Betal97}, and S.~Cassisi (1999, private
communication). The tracks with 3.5 solar metallicity are taken from
\citet{Saletal00}. Lick indices are computed by adopting the fitting
functions of \citet{Woretal94}. A \citet{Salpeter55} IMF is adopted.

\paragraph*{\boldmath \aFe\ enhancement.---}The 
$\alpha$/Fe enhanced mixtures are produced by increasing the
abundances of the $\alpha$-group-elements N, O, Mg, Ca, Na, Ne, S, Si,
Ti, and by decreasing the Fe-peak element (Cr, Mn, Fe, Co, Ni, Cu, and
Zn) abundances, such that total metallicity is conserved (see T00).
In additional models, the elements nitrogen and calcium are detached
from the $\alpha$-group, and Lick indices for various \aN\ and \aCa\
ratios are computed.  The effect from these element abundance changes
on the Lick indices are taken from TB95. These authors computed model
atmospheres and synthetic spectra for the three evolutionary phases
dwarfs, turnoff, and giants of a 5~Gyr old isochrone with solar
metallicity. They double in turn the abundances of the dominant
$\alpha$- and Fe-peak elements, and determine for each phase
separately the resulting index changes. The TB95 fractional changes
are incorporated in the models using an extension of the method
introduced by T00.

\paragraph*{Three evolutionary phases.---}We 
compute the final SSP models in the following way. The basic SSP model
is divided in the three evolutionary phases as defined in TB95. The
standard Lick indices are computed for each phase separately and
modified according to the desired element abundance ratio using the
index responses from TB95. The final index is the flux-weighted sum
over the three phases.

\paragraph*{\boldmath \aFe\ bias.---} Most
importantly, we take into account that any stellar population model
being based on stellar libraries constructed from Milky Way stars
reflects the chemical enrichment history of the Milky Way. This
implies that standard model Lick indices are biased toward super-solar
\aFe\ ratios at sub-solar metallicities. Accounting for this bias, the
models presented here have well-defined \aFe\ ratios at all
metallicities.


\section{Calibration on globular clusters}
\label{sec:calibration}
Globular clusters are the observed counterparts of theoretical simple
stellar populations, because their stars are coeval and have all the
same chemical composition.  Therefore, globular clusters are the ideal
targets for calibration purposes.  By using a new set of globular
cluster data \citep{Puzetal02}, the metallicities of which extend up
to solar, \citet{Maretal02} check the 21 Lick indices of the base
model. The result is that above $[\ZH]\ga -1$ the standard model is
not able to reproduce the data because effects from \aFe\ ratios are
not included. To solve this problem, we construct the present \aFe\
enhanced SSP models. In this section we present their calibration with
globular cluster data.

Galactic globular clusters in the halo and in the bulge appear coeval
independent of their metallicities \citep{Ortetal95,Roetal99,Petal00}.
Ages between 9 and 14~Gyr which are derived from color-magnitude
diagrams \citep{Van00}. Therefore the calibration performed here tests
the models with old ages.  For a calibration at younger ages with the
globular clusters of the Large Magellanic Cloud see \citet*{BHS02}.

\subsection{\boldmath\aFe\ ratios}
\begin{figure*}
\psfig{file=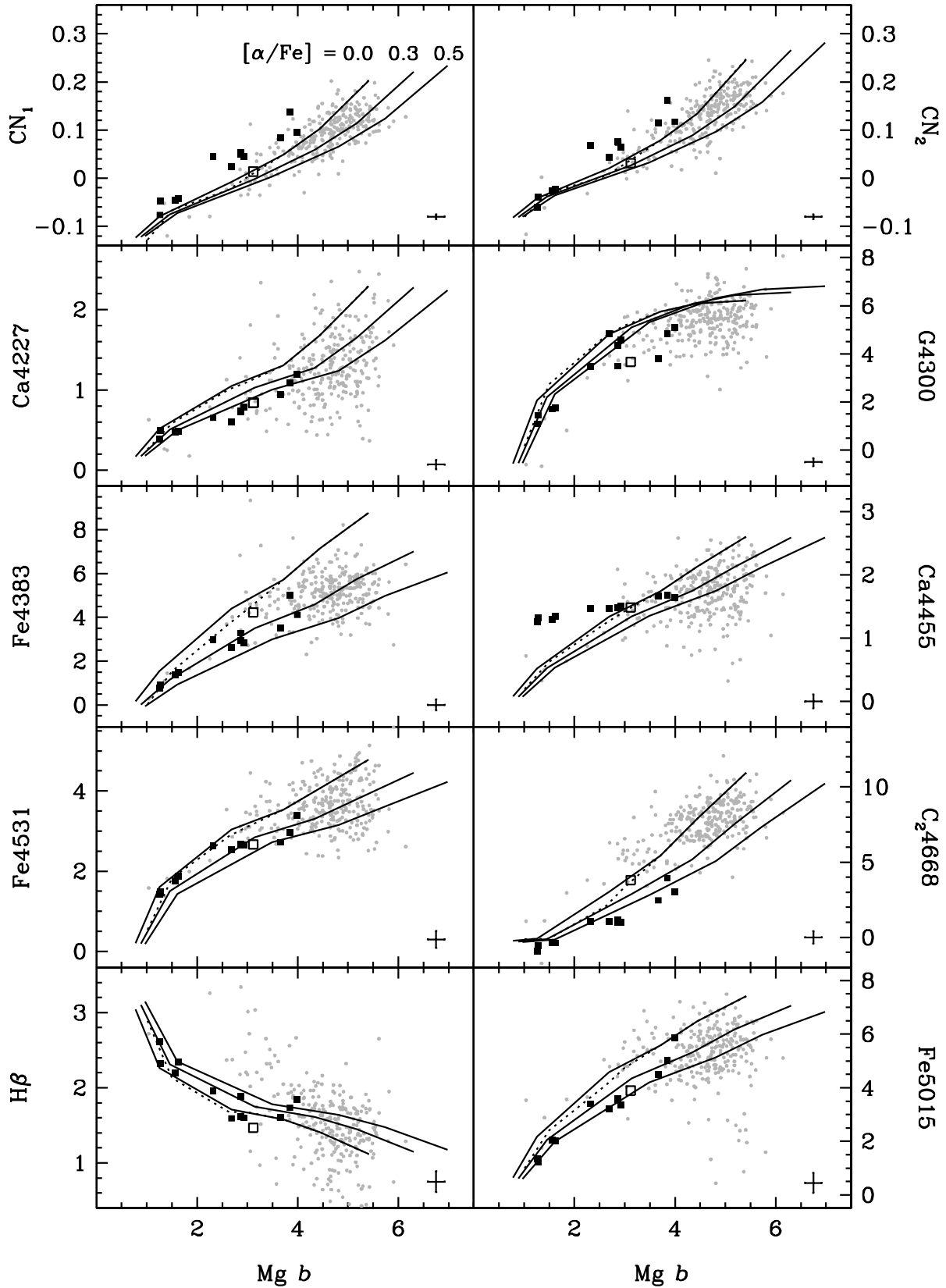,width=0.98\linewidth}
\caption{\Mgb\ index versus the other 20 Lick indices. Solid lines are
the models of this paper with constant age (12~Gyr), constant \aFe,
and for the metallicity range $-2.25\leq [\ZH]\leq 0.67$. Three models
with $[\aFe]=0.0,\ 0.3,\ 0.5$ are shown. Models with solar abundance
ratios ($[\aFe]=0.0$) and models with $[\aFe]=0.5$ are those with the
lowest and highest \Mgb\ indices, respectively. The dotted lines are
our base SSP models \citep{Ma98}. Filled squares are globular cluster
data, the open square is the integrated Bulge light from
\citet{Puzetal02}, small grey dots are the Lick data of giant
elliptical galaxies from \citet{Traetal98}. Errorbars indicate typical
errors of the globular cluster data.}
\label{fig:allindices}
\end{figure*}
\begin{figure*}
\psfig{file=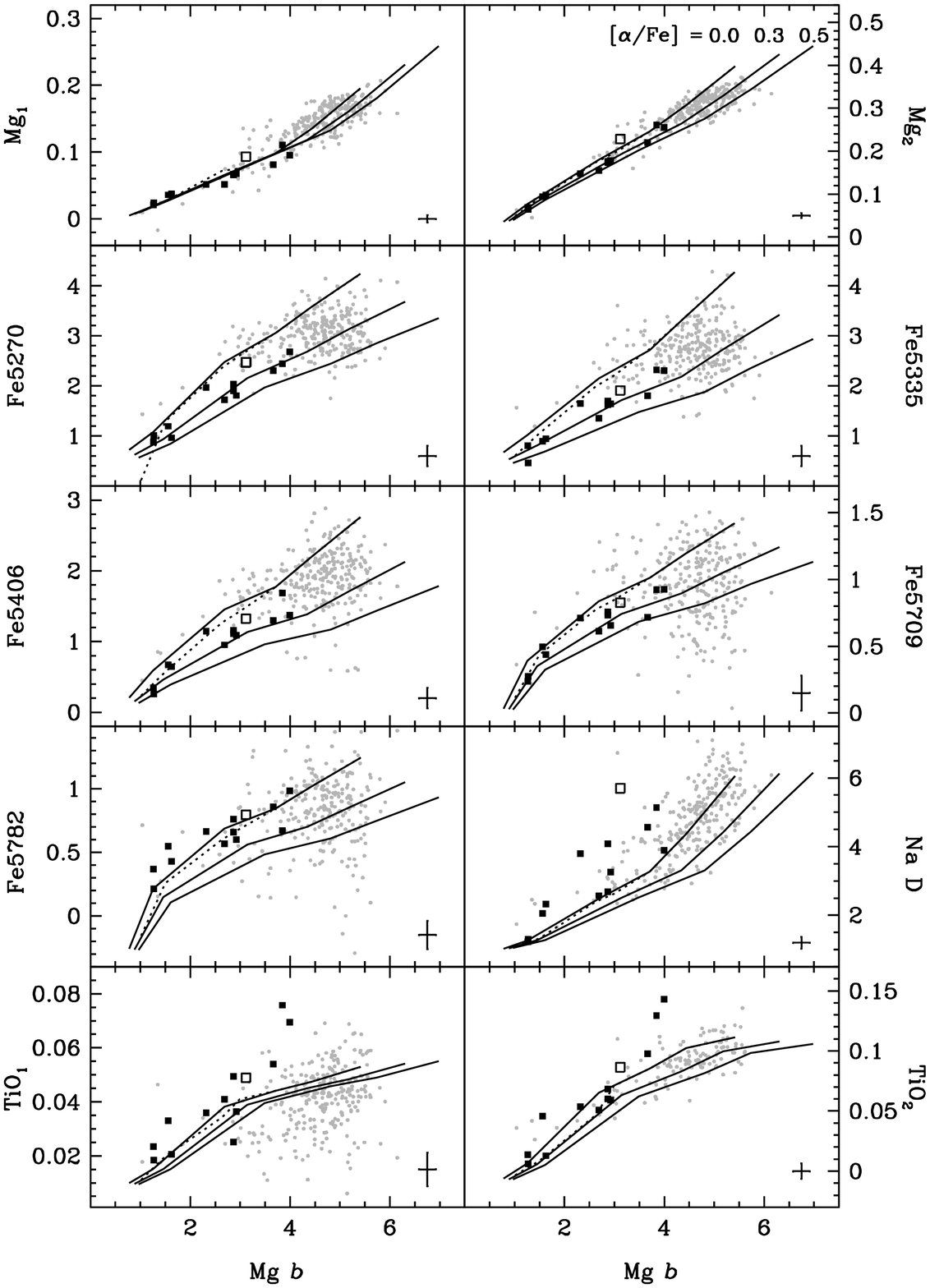,width=0.98\linewidth}
\contcaption{}
\end{figure*}

From high-resolution stellar spectroscopy it is known that globular
clusters in both the halo and the Bulge of our Galaxy are \aFe\
enhanced with the typical values $[\alpha/{\rm Fe}]\approx 0.3\pm
0.1$~dex (\citealt{Baretal99}, \citealt{Cohetal99},
\citealt{Caretal01}, \citealt{Coeetal01}, \citealt*{ORC02}, and
compilations by \citealt{Ca96}, and \citealt{SC96}). Hence, the
observed Lick indices of globular clusters should be matched by SSP
models with $t\approx 12$~Gyr (see above) and $[\aFe]\approx 0.3$.

The comparison is shown in Fig.~\ref{fig:allindices}, in which we plot
the \Mgb\ index versus the other 20 Lick indices. The indices are
arranged according to increasing wavelength. Filled squares are
globular cluster data \citep{Puzetal02}, solid lines are the new SSP
models presented here with constant age (12~Gyr) and constant \aFe\
ratio. All models cover the metallicity range $-2.25\leq [\ZH]\leq
0.67$. The three lines are models with the abundance ratios
$[\aFe]=0.0,\ 0.3,\ 0.5$~dex. Models with solar abundance ratios
($[\aFe]=0.0$) and models with $[\aFe]=0.5$ are those with the lowest
and highest \Mgb\ indices, respectively. The dotted lines are the base
SSP model.  For well calibrated indices, the squares (globular cluster
data) should scatter about the middle line (model with
$[\aFe]=0.3$). The open square is the integrated Bulge light
\citep{Puzetal02}, small grey symbols are the Lick data of giant
elliptical galaxies from \citet{Traetal98}.

\subsubsection{The \aFe\ bias in standard models}
The standard models are shown as dotted lines in
Fig.~\ref{fig:allindices}.  As their \aFe\ bias is $[\aFe]\approx 0.3$
at the lowest metallicity (see Table~\ref{tab:bias}), this biased
model (dotted line) deviates clearly from the $[\aFe]=0$ model and
coincides with the $[\aFe]=0.3$ model (middle solid line). Indeed, as
shown in \citet{Maretal02}, because of the \aFe\ bias the base model
matches the Lick data of the metal-poor globular clusters.  At solar
metallicity and above, there is no bias in the standard model
(Table~\ref{tab:bias}), so that the dotted lines and the $[\aFe]=0.0$
model (solid line with the lowest \Mgb) are indistinguishable for
$[\ZH]\geq 0$. This pattern is present throughout all panels in
Fig.~\ref{fig:allindices}.

\subsubsection{Sensitivity to Fe-peak elements}
As emphasized in T00, and explained at the beginning of
Section~\ref{sec:abundancevar}, the enhancement of the \aFe\ ratio at
fixed metallicity is produced essentially by a depletion of the Fe
abundance (see also \citealt{BGM92}), and only by a slight increase of
the $\alpha$-element abundances. The reason is that Fe-peak elements
are by far less abundant than oxygen and the other
$\alpha$-elements. As a consequence, only indices that are sensitive
to Fe-peak element abundance variations respond to \aFe\ ratio changes
at fixed metallicity. \Mgb, for instance, anti-correlates with Fe
abundance (TB95), and therefore increases with increasing \aFe.

Ca4227, instead, correlates with both Ca and Fe abundance (TB95). As
the depletion of iron dominates the \aFe\ enhanced model, Ca4227
actually decreases with increasing \aFe, although the element calcium
belongs to the enhanced group in our models. The indices \CNone\ and
\CNtwo\ correlate with \aFe, also mainly because of their sensitivity
to Fe-peak element abundances.

\subsubsection{Discussion on individual indices}
\label{sec:indexdiscussion}
\begin{figure*}
\psfig{file=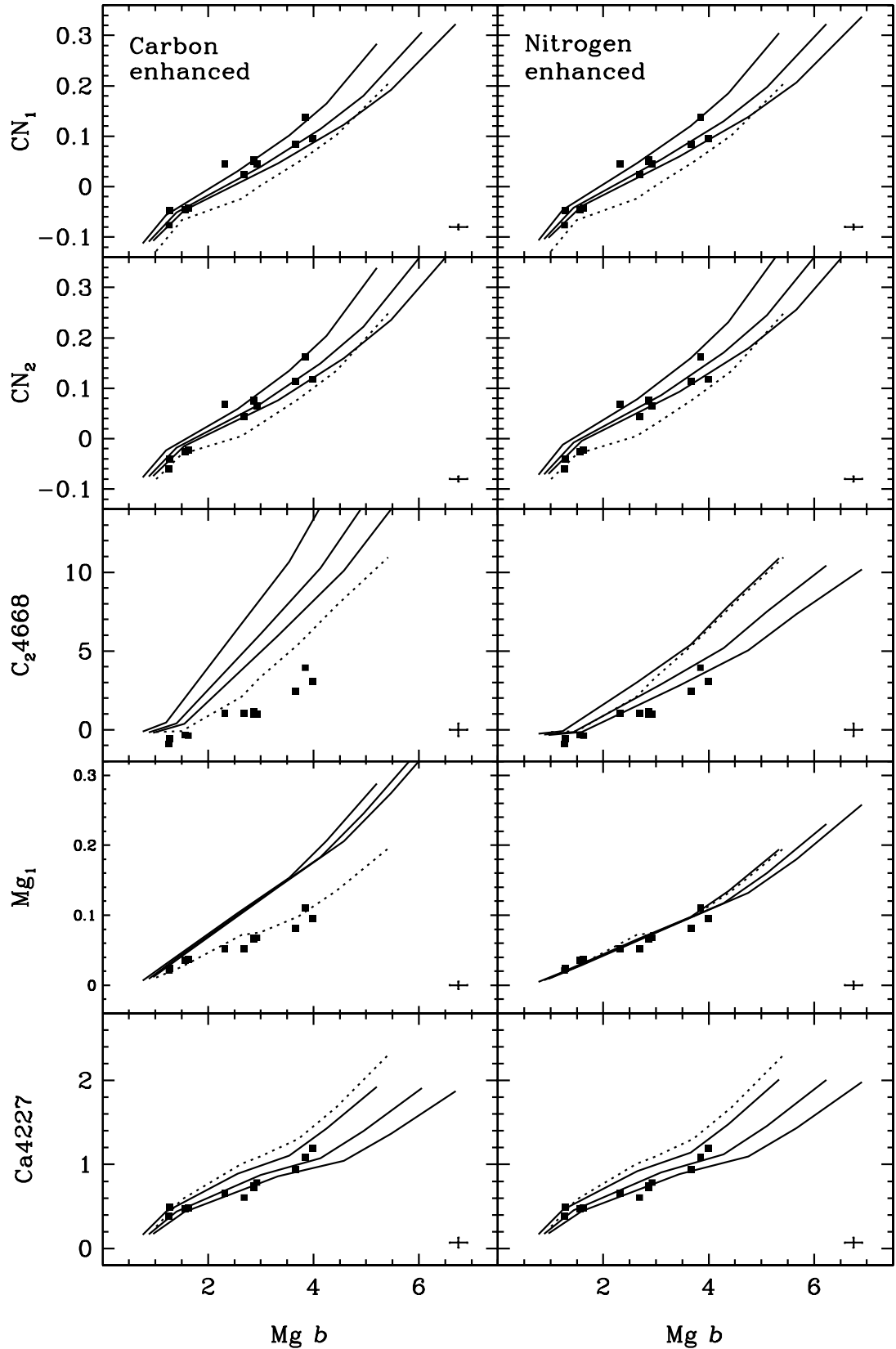,width=0.98\linewidth}
\caption{Models in which carbon is enhanced by 0.1 dex (left-hand
panels) and in which nitrogen is enhanced by 0.5 dex (right-hand
panels). Only carbon and nitrogen sensitive indices are
shown. Line-styles and symbols like in Fig.~\ref{fig:allindices}.}
\label{fig:cn}
\end{figure*}
\CNone,\CNtwo.---The models clearly predict too low index values at
all metallicities, in particular for \CNone, although both indices
respond positively to changes in the \aFe\ ratios. Such relatively
strong CN absorption features have also been observed for
extragalactic globular clusters in M~31 \citep{Buretal84} and NGC~3115
\citep{Kunetal02}. As shown in \citet{Maretal02}, the mismatch between
models and data is already present in the standard SSP model (dotted
lines), thus cannot be attributed to a failure of the TB95 index
responses. Moreover, the integrated light of the galactic Bulge (open
squares) does not have such strong CN absorption features.

It has been suggested by \citet*{DGC83} and \citet{Renzini83}, that
stars in globular clusters may accrete carbon and/or nitrogen enriched
ejecta from the surrounding AGB stars \citep{RV81}. Indeed \CNone\ and
\CNtwo\ are very sensitive to carbon and to nitrogen abundances
(TB95).  To test this option quantitatively, we computed additional
models enhancing separately the abundances of C and N. The resulting
models for the indices \CNone, \CNtwo, Ca4227, \FeC, and \Mgone\ are
shown in Fig.~\ref{fig:cn}. The remaining Lick indices (except G4300)
are not affected by carbon and nitrogen abundance changes.

An increase of the carbon abundance by 30 per cent is sufficient to
obtain an excellent match of the \CNone\ and \CNtwo\ data (left-hand,
top panels in Fig.~\ref{fig:cn}). However, also \FeC\ and \Mgone\
strongly correlate with the element carbon, so that these indices are
not reproduced any more by the models with enhanced carbon abundance
(left-hand middle panels in Fig.~\ref{fig:cn}). It is not possible to
match simultaneously the four indices \CNone, \CNtwo, \FeC, and
\Mgone.  Assuming that the index responses by TB95 are correct, we can
rule out that a significant enhancement of the carbon abundance
produces the \CNone\ and \CNtwo\ indices of globular clusters
\citep[see also][]{Wor98}.

Increasing the nitrogen abundance by a factor 3 with respect to the
$\alpha$-elements ($[\aN]=-0.5$), instead, the indices \CNone\ and
\CNtwo\ are well reproduced, without destroying the match to the other
indices, as the latter are almost insensitive to nitrogen abundance
(right-hand panels in Fig.~\ref{fig:cn},
Table~\ref{tab:N-models}). Based on the TB95 calculations, already
\citet{Wor98} argued that an enhancement of nitrogen rather than
carbon is required to explain the CN-absorption in globular cluster
stars. The model calculations of this paper now allow us to quantify
this conclusion. We show that nitrogen needs to be enhanced in
globular cluster stars by 0.5~dex (factor~3). This is consistent with
the results from \citet{ORC02}, who find nitrogen to be enhanced by a
factor $\approx 2$ in stars in NGC~6553, while carbon is not
enhanced. It should be emphasized again that the CN features of the
Bulge light, instead, are perfectly reproduced by the models without
extra-enhancement of nitrogen.

\medskip
Ca4227.--The absorption index Ca4227 is predicted slightly too high by
the models. Interestingly, Ca4227 is the only index besides \CNone\
and \CNtwo\ which is considerably affected by changes of carbon and
nitrogen abundances. Its line-strength is anti-correlated with CN
abundances (TB95), so that the poor match of the globular cluster data
in Fig.~\ref{fig:allindices} is improved with the CN-enhanced models
shown in the bottom panels of Fig.~\ref{fig:cn}.  Taking the
enhancement of nitrogen into account, our \aFe\ enhanced models with
$[\CaFe]=[\aFe]$, provide a good fit to the globular cluster data,
including the metal-rich clusters NGC~6528 and NGC~6553. This
indicates that the latter do not exhibit any anomalies in calcium
abundance, i.e.\ their Ca/Fe ratio are enhanced like the other
$\alpha$-element/Fe ratios. This result agrees with element abundances
derived for single stars in these clusters \citep{Caretal01,ORC02},
even though the latter are still controversial. Barbuy and coworkers
(B.\ Barbuy, 2002, private communication) measure $[{\rm Mg}/{\rm
Fe}]=0.3$ but $[\CaFe]=0.0$ in a clump star of NGC~6528, which would
imply a Ca underabundance in that cluster.

Motivated by the realization that elliptical galaxies have weaker
Ca4227 indices than predicted by standard SSP models
\citep{Vazetal97,Traetal98}, we additionally computed models of Ca4227
with variable \aCa\ ratios (Table~\ref{tab:Ca-models}). The
application of these models and the derivation of \aCa\ ratios for
elliptical galaxies is presented in an accompanying paper (D.~Thomas
et al., in preparation).

\medskip
G4300.--This index is mainly sensitive to carbon and oxygen
abundances, only little to Fe abundance (TB95). It responds therefore
only marginally to the \aFe\ ratio changes, as enhanced \aFe\ ratios
are not caused by $\alpha$-element enhancement, but by Fe
reduction. At high metallicities, it is even less sensitive to total
metallicity than \Hb\ (see also TB95).  The calibration of the models
with the globular cluster data is not convincing.

\medskip
Fe4383.--This index is very sensitive to Fe abundance, hence \aFe\
ratios. The models provide an excellent fit to the data, so that
Fe4383 is well calibrated and represents a promising (blue)
alternative to the classic indices Fe5270 and Fe5335.

\medskip
Ca4455.--As emphasized by TB95, despite its name Ca4455 is insensitive
to Ca abundance, while Fe and Cr, both elements of the depressed
group, are the dominant contributors to this index. As Ca4455 is
correlated to Cr, but anti-correlated to Fe abundance, however, it
responds hardly to \aFe\ ratios changes. The globular cluster data and
the model predictions are not compatible as already shown in case of
the base model \citep{Maretal02}. It is more likely that this mismatch
originates from an offset of the globular cluster data from the Lick
system \citep[see][]{Maretal02}, because these data seem also
incompatible with the Lick galaxy data of \citet[][small grey
symbols]{Traetal98}. The index Ca4455 is therefore not a useful
abundance indicator.

\medskip
Fe4531.--This index is reasonably well calibrated, but it is less
sensitive to Fe abundance and \aFe\ ratios than Fe4383.

\medskip
\FeC.--Formerly called Fe4668, this index has been renamed, because it
is most sensitive to carbon abundance (TB95). As carbon is kept fixed
in our models, \FeC\ decreases only slightly with increasing
\aFe. Owing to the poor match between models and globular cluster
data, this index is not well suited for element abundance studies.
Assigning carbon to the depressed group, does not improve the match
between data and models, and provokes inconsistencies with the
otherwise well calibrated indices \Mgone, \CNone, and \CNtwo.

\medskip
\Hb.--The Balmer absorption index is well calibrated. It increases
mildly with increasing \aFe.  The increase of Balmer absorption in
globular clusters with decreasing metallicity is very well reproduced
by our models \citep[see also][]{MT00}. This strong Balmer absorption
at old ages and low metallicities stems from the development of warm
horizontal branches owing to mass loss on the Red Giant Branch. As
shown in \citet{Maretal02}, the slightly lower \Hb\ values of the two
globular clusters at intermediate metallicity ($\Mgb\approx 2.8$) can
be easily reproduced by models with reduced mass loss along the red
giant branch, in good agreement with their observed red horizontal
branch morphologies. The models used here have blue horizontal
branches at metallicities below $[\ZH]\sim -1$.

\medskip
Fe5015.--Although TB95 produce only a poor fit of this index, our
models are in good agreement with the globular cluster data. As Fe5015
is only little sensitive to \aFe\ ratios, however, it is less
recommendable for abundance ratio studies than Fe4383.

\medskip
\Mgone,\Mgtwo.--It is very reassuring that all three Mg-indices
\Mgone, \Mgtwo, and \Mgb\ respond very similarly to \aFe\ ratio
changes. The models plotted in Fig.~\ref{fig:allindices} are therefore
highly degenerate. The globular cluster data are well
reproduced. Among the three indices, \Mgtwo\ turns out to be least,
\Mgb\ to be most sensitive to \aFe.

\medskip
Fe5270,Fe5335.--These are the classical indicators for Fe
abundance. They are perfectly matched by our models. Fe5335 is
somewhat more sensitive to \aFe.

\medskip
Fe5406.--This index is very similar to Fe5270 and Fe5335. Although the
\aFe\ enhanced models predict slightly lower Fe5406 values than
suggested by the data, this index is still reasonably well calibrated.

\medskip
Fe5709.--The match between models and data is excellent. However,
Fe5709 responds less strongly to \aFe\ because of its weaker
sensitivity to Fe abundance (TB95).

\medskip
Fe5782.--For this index, the match between models and globular cluster
data is very poor. Different from the situation of Ca4455, it is hard
to assess if the globular cluster data are compatible with the Lick
galaxy measurements of \citet{Traetal98}. The standard SSP models
(dotted line), which are perfectly compatible with the \citet{Wo94}
models, clearly predict too low Fe5782 indices. A miscalibration of
the fitting function therefore may also be a possible explanation for
the mismatch between models and data \citep{Maretal02}.

\medskip
Na~D.--Similar to Fe5782, the standard SSP model gives lower index
values than suggested by the observational data, in accordance with
the \citet{Wo94} models. An unrealisticly large sensitivity of the
index to the \aFe\ ratio would be required to lift the models on the
data. Most likely, the strong NaD absorption is caused by Na
absorption in interstellar material of the Galactic disc. Indeed,
there is the clear trend that the clusters in the sample that are
closer to the Galactic plane have higher NaD relative to their \Mgb\
indices.  This high sensitivity of the NaD index to interstellar
absorption severely hampers its usefulness for stellar population
studies.

\medskip
\TiOone,\TiOtwo.--Both indices appear badly calibrated, although the
agreement between models and data for \TiOtwo\ is still acceptable. As
discussed in \citet{Puzetal02}, the most metal-rich clusters NGC~6528
and NGC~6553 show very strong \TiOtwo\ absorption because of their
extremely cool Red Giant Branches, in accordance with the strong
bending observed in color-magnitude diagrams \citep*{OBB91,CS95}.

\subsubsection{Measuring \aFe\ from \Mgb/\Fe}
\begin{figure}
\psfig{file=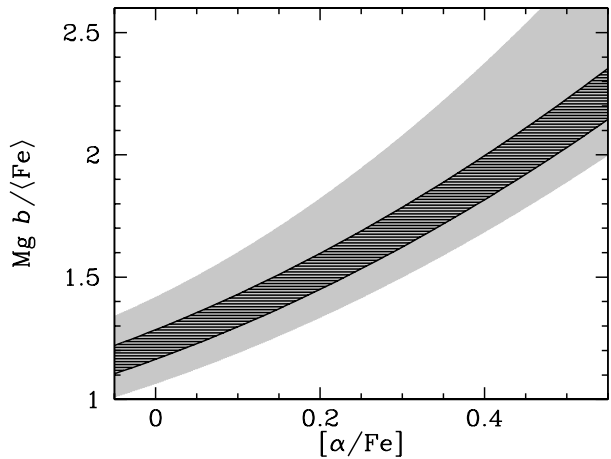,width=\linewidth}
\caption{Lick index ratio \Mgb/\Fe\ as a function of element abundance
ratio \aFe. The dark hashed area are models covering metallicities
$-1.35\leq [\ZH]\leq 0.35$ and ages from 8 to 15~Gyr. The light-grey
area are models with the same metallicities but the larger age range
$3\leq t\leq 15$~Gyr. }
\label{fig:afemgbfe}
\end{figure}
The models provided here allow for the derivation of \aFe\ element
ratios from the Lick indices like \Mgb\ and \Fe.  In the following, we
show how the index ratio \Mgb/\Fe\ can be used to obtain an estimate
of the element ratio \aFe. For this purpose, in
Fig.~\ref{fig:afemgbfe} we plot \Mgb/\Fe\ as a function of \aFe. As
\Mgb/\Fe\ depends not only on \aFe\ but also on age and metallicity,
we show the areas covered by models with a range in ages and
metallicities.  The dark shaded area are models covering metallicities
$-1.35\leq [\ZH]\leq 0.35$ and ages from 8 to 15~Gyr. The light-grey
area are models with the same metallicities but the larger age range
$3\leq t\leq 15$~Gyr.

For old populations (ages above $\sim 8$~Gyr), the relation between
\Mgb/\Fe\ and \aFe\ is reasonably well defined independent of age and
metallicity. In this case, Fig.~\ref{fig:afemgbfe} allows to read off
directly the \aFe\ element ratio $\pm 0.04$~dex from the measured
index ratio \Mgb/\Fe.

\subsubsection{\aFe\ enhanced stellar evolutionary tracks}
In the models presented here, solar-scaled stellar tracks are adopted.
The impact on Lick indices due to \aFe\ enhancement is accounted for
through a modification of the stellar absorption line-strengths (see
Section~\ref{sec:construction}). A fully self-consistent \aFe\
enhanced SSP model should in principle use \aFe\ enhanced stellar
evolutionary tracks, because the element abundance variations in a
star affect also the star's evolution and the opacities in the stellar
atmosphere, hence the effective temperature.  We are planning in
future to compute SSP models with \aFe\ enhanced tracks that are
based---for reasons of self-consistency---on the same input tracks
\citep{CCC97} as the present models.

To assess the impact of \aFe\ in the stellar evolutionary tracks,
\citet{Maretal02} compute standard models with the solar-scaled and
the \aFe\ enhanced tracks from \citet{Saletal00}. As the \aFe\
enhanced tracks are hotter than the solar-scaled ones
\citep{Saletal00}, their inclusion in the stellar population model
leads to slightly weaker metallic indices (i.e.\ \Mgb, \Fe\ etc.) and
stronger Balmer line indices (\Hb) for the same age and metallicity
\citep[Fig.~5 in][]{Maretal02}. The decrease of \Mgb\ and \Fe\ are
comparable, so that the additional inclusion of \aFe\ enhanced tracks
has only a minor effect on the \Mgb-\Fe\ plane, and therefore has no
significant impact on the derivation of \aFe\ ratios. This issue is
explored in detail in an accompanying paper \citep{TM02}.

\subsubsection{Summary}
To briefly summarize, the classical indices \Mgone, \Mgtwo, \Mgb\ and
the blue indices \CNone\ and \CNtwo\ increase with increasing \aFe\
ratio, in particular the latter owing to an anti-correlation with Fe
abundance. With the caveat that \CNone\ and \CNtwo\ are very sensitive
to C and N abundances, these two can be regarded complementary to the
indices Mg$_1$, Mg$_2$, \Mgb.  Besides the intensively studied iron
indices Fe5270 and Fe5335, the indices Fe4383, Fe4531, Fe5015, and
Fe5709 are good representatives of Fe-peak element abundances.  The
indices G4300, Ca4455, \FeC, Fe5782, Na~D, \TiOone, instead, are
poorly calibrated and do not provide valuable information on abundance
ratios. Ca4227, \Hb, Fe5406 and \TiOtwo\ cannot be assigned to any of
these three categories.

\Hb\ is only little sensitive to element abundance variations, and is
well calibrated. Ca4227 is mainly sensitive to Ca and N abundances.
This index, and the indices \CNone\ and \CNtwo, require an additional
enhancement of nitrogen abundance with respect to the other
$\alpha$-elements by a factor 3 ($[\aN]=-0.5$), in order to fit the
globular cluster data.

Concluding, the combination of the blue indices \CNone, \CNtwo\ and
Fe4383 may be best suited to estimate \aFe\ ratios of objects at
redshifts $z\sim 1$.

\subsection{Metallicities}

\subsubsection{The globular cluster metallicity scale}
In this section we compare the total metallicities [\ZH] derived here
for the galactic globular cluster sample with the metallicities given
in the \citet{Ha96} catalog, which are based on the \citet[][hereafter
ZW84]{ZW84} scale.  We add the data of the globular cluster 47~Tuc
from \citet{Maretal02}\footnote{The original spectrum comes from
\citet*{CGP95}.  The Lick indices have been measured in the
\citet{Woretal94} system by \citet{Maretal02} on the spectrum provided
by S.~Covino (2002, private communication).}.  For the metal-rich
bulge cluster NGC~6553 we adopt the more recent measurements of single
star abundances by \citet{Baretal99}, who find---in good agreement
with \citet{ORC02}---solar total metallicity.
\begin{figure}
\psfig{file=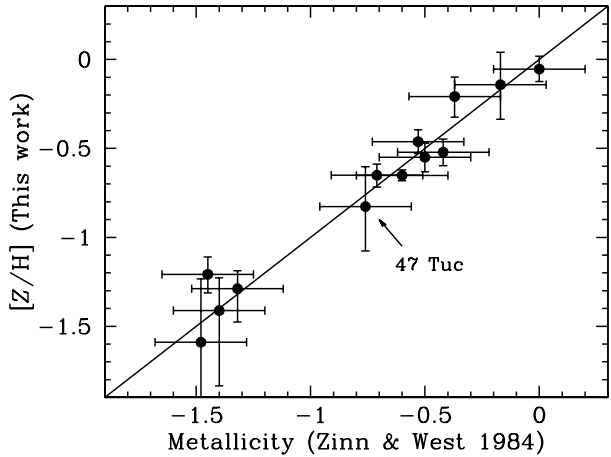,width=\linewidth}
\caption{\citet{ZW84} metallicity \citep{Ha96} vs.\ the total
metallicity [\ZH] derived in this paper for galactic globular clusters
from \citet{Puzetal02}.  [\ZH] is determined from the indices \Mgb\
and Fe5270, using our SSP models with fixed age $t=12$~Gyr. We assume
an error of 0.2~dex for the ZW84 metallicities.}
\label{fig:zscale}
\end{figure}

ZW84 metallicities are usually referred to as [Fe/H]. As we actually
aim at calibrating our SSP models, in which we distinguish between
total metallicity [\ZH] and iron abundance [\FeH], at this point it is
crucial to understand, if the ZW84 metallicity scale traces total
metallicity or iron abundance.  It is important to remember, that the
ZW84 scale at $[\FeH]>-1.5$ is based on measurements by
\citet{Cohen83} of the pseudo-equivalent widths for the Mg triplet
near 5175~\AA, and the 5270 and 5206~\AA\ Fe blends. She derives the
metallicity from averaging the Mg and Fe equivalent widths. These
metallicities are then used by ZW84 to set up a globular cluster
metallicity scale.  The average of Mg and Fe line kills information on
\aFe\ ratios and is likely to be close to the total metallicity.  This
strongly suggests that the ZW84 metallicity scale traces total
metallicity rather than iron abundance at $[\ZH]\ga -1.5$.  We suggest
that the ZW84 metallicities should better be written as [\ZH] instead
of [Fe/H], in order to avoid confusion.

We determine [\ZH] from the Lick indices \Mgb\ and Fe5270, using our
SSP models with fixed age $t=12$~Gyr. Note that using alternatively
the indices \Mgtwo\ and/or Fe5335 yields perfectly consistent results.
Fig.~\ref{fig:zscale} shows that our derived total metallicities [\ZH]
are in excellent agreement with the ZW84 metallicity scale.  This
result is highly reassuring but not unexpected, as we derive
metallicities from exactly the same lines (Mg triplet at 5157~\AA\ and
Fe blend at 5270~\AA) as \citet{Cohen83}, even though from integrated
light and at the much lower Lick resolution ($\sim 8$~\AA).

The iron abundance [Fe/H] can be obtained through the following
scaling (see Section~\ref{sec:abundancevar},
Eqn.~\ref{eqn:Fescaling}),
\[
[\FeH]=[\ZH]-0.94\ [\aFe]
\]
and is lower than [\ZH] for super-solar \aFe\ ratios. The [\ZH] and
[\aFe] ratios derived here for the globular cluster sample are listed
in Table~\ref{tab:abundances}. The iron abundances of all the
globulars considered here are lower than the total metallicity by
typically $\sim 0.3$ dex. Our measured [Fe/H] are therefore
systematically lower by this value than the ZW84 metallicities,
reinforcing the interpretation that the ZW84 scale implies total
metallicity.
\begin{table}
\caption{Derived element abundances}
\begin{tabular}{ccc}
\hline\hline
Name      &   [\ZH] & [\aFe] \\
\hline
 NGC 6528  &    $-0.14\ \pm 0.19$   &   $0.31\ \pm 0.32$ \\
 NGC 6553  &    $-0.05\ \pm 0.07$   &   $0.23\ \pm 0.12$ \\
 NGC 5927  &    $-0.21\ \pm 0.11$   &   $0.35\ \pm 0.21$ \\
 NGC 6388  &    $-0.65\ \pm 0.03$   &   $0.08\ \pm 0.04$ \\
 NGC 6624  &    $-0.52\ \pm 0.08$   &   $0.33\ \pm 0.09$ \\
 NGC 6218  &    $-1.59\ \pm 0.43$   &   $0.24\ \pm 0.45$ \\
 NGC 6441  &    $-0.46\ \pm 0.07$   &   $0.26\ \pm 0.07$ \\
 NGC 6626  &    $-1.21\ \pm 0.10$   &   $0.14\ \pm 0.21$ \\
 NGC 6284  &    $-1.29\ \pm 0.15$   &   $0.38\ \pm 0.24$ \\
 NGC 6356  &    $-0.55\ \pm 0.08$   &   $0.40\ \pm 0.10$ \\
 NGC 6637  &    $-0.65\ \pm 0.07$   &   $0.35\ \pm 0.08$ \\
 NGC 6981  &    $-1.41\ \pm 0.30$   &   $0.09\ \pm 0.35$ \\
 47 Tuc    &    $-0.83\ \pm 0.23$   &   $0.19\ \pm 0.35$ \\
\hline
\end{tabular}
\label{tab:abundances}
\end{table}

Based on high resolution spectroscopy of individual stars in globular
clusters, \citet{CG97} suggest a revision of the classical ZW84 scale
to essentially higher metallicities. The shift is of the order
0.2~dex, which leads to a clear disagreement between the metallicities
derived here from Lick indices and the metallicity scale of
\citet{CG97}. The same discrepancy emerges from considering integrated
colors of globular clusters.  \citet{Maraston00} show that SSP models
using reasonable ages ($\ga 9$~Gyr) and the \citet{CG97} metallicities
of globular clusters predict \BV\ colors that are too red compared to
the observed values. With the ZW84 metallicities, instead, excellent
agreement between SSP model prediction and observation is
found. \citet{CC02} come to the same conclusion comparing isochrones
with observed color-magnitude diagrams of globular clusters.

Interestingly, in contrast to \citet{Cohen83}, \citet{CG97} measure
'metallicity' exclusively from Fe line features, so that their values
should rather be iron abundances than total metallicities. Taking this
into account, the discrepancy to the metallicities derived from
isochrones and SSP models gets even more severe \citep[see Fig. 8
in][]{Maretal02}. \citet{Barbuy00} suggests that \citet{CG97} may tend
to overestimate element abundances and metallicities because of a
hotter temperature scale.

A comparison between the metallicities of globular clusters derived
from Lick indices and the ZW84 metallicity scale has also been carried
out by \citet*{CBR98}. They used \citet{Wo94}'s SSP models to fit
simultaneously by $\chi^2$ minimization the indices \Mgb, NaD, Fe5270,
and Fe5335 measured for globular clusters including also the
metal-rich bulge cluster NGC~6528. The authors did not find a
satisfying consistency between their derived metallicities and the
ZW84 scale. They had to introduce a scaling between ZW84 and their
derived metallicities, as they over-predicted metallicities for the
more metal-rich clusters. This may partly be due to the higher Fe
indices measured \citep[see][]{Puzetal02}, partly due to the fact that
\citet{Wo94}'s SSP models do not include \aFe\ abundance effects.

Concluding, we would like to emphasize again that metallicity
determinations from colors and Lick indices with the models of this
paper are both in excellent agreement with the ZW84 metallicity scale.
This highly encouraging self-consistency is also found by
\citet{Kunetal02}, who derive---with the models of this paper---the
same metallicities of globular clusters in the elliptical galaxy
NGC~3115 from the Lick indices \Mgb\ and \Fe, and \VI\ colors.

\subsubsection{Brodie \& Huchra's metallicity calibration}
\begin{figure}
\psfig{file=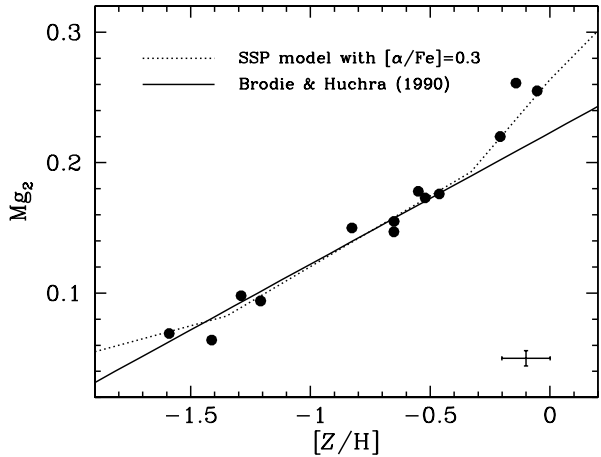,width=\linewidth}
\caption{Total metallicity [\ZH] versus \Mgtwo\ index. The circles are
galactic globular clusters \citep{Puzetal02}. [\ZH] is determined from
the indices \Mgb\ and Fe5270, using the SSP models of this paper with
fixed age $t=12$~Gyr (see also Fig.~\ref{fig:zscale}). The solid line
is the metallicity calibration by \citet{BH90}. The dotted line is the
SSP model of this paper with fixed age $t=12$~Gyr and $[\aFe]=0.3$.}
\label{fig:brodie}
\end{figure}

Measuring line indices on galactic and M31 globular cluster spectra,
\citet{BH90} derived a linear correlation between \Mgtwo\ index and
ZW84 metallicity. For a comparison of our derived metallicities [\ZH]
with this calibration, in Fig.~\ref{fig:brodie} we show [\ZH] of the
globular cluster sample versus \Mgtwo, and over-plot as a solid line
the linear relation derived by \citet{BH90}. The dotted line is our
SSP model with age $t=12$~Gyr and $[\aFe]=0.3$.

Although reassuring, the excellent agreement does certainly not come
as a surprise, as \citet{BH90} derived their fit on the basis of ZW84
metallicities, which are perfectly consistent with our [\ZH] values
(Fig.~\ref{fig:zscale}). More interesting are the two most metal-rich
clusters NGC~6528 and NGC~6553, which exhibit significantly stronger
\Mgtwo\ than predicted by the \citet{BH90} calibration. These clusters
have almost solar metallicities, while the \citet{BH90} calibration is
based on globular clusters with $[\ZH]<-0.5$. A non-linear increase of
\Mgtwo\ with increasing total metallicity at metallicities
$[\ZH]>-0.5$, instead, is suggested by the data and consistently
predicted by our SSP models (dotted line).

\subsubsection{Tracing metallicity independent of \aFe}
The new SSP models with variable \aFe\ ratio allow for an unambiguous
derivation of total metallicity and \aFe\ ratio simultaneously, free
from any \aFe\ bias. Still, it would be useful to find an index that
is mainly a tracer of total metallicity independent of the \aFe\
ratio.  \citet{G93} suggested that averaging Mg and Fe indices may
yield such a metallicity indicator, and defined the index
\[
[{\rm MgFe}]=\sqrt{\Mgb\cdot\Fe}
\]
with 
\[
\Fe=\frac{1}{2}\ ({\rm Fe5270}+{\rm Fe5335})\ .
\]

\begin{figure}
\psfig{file=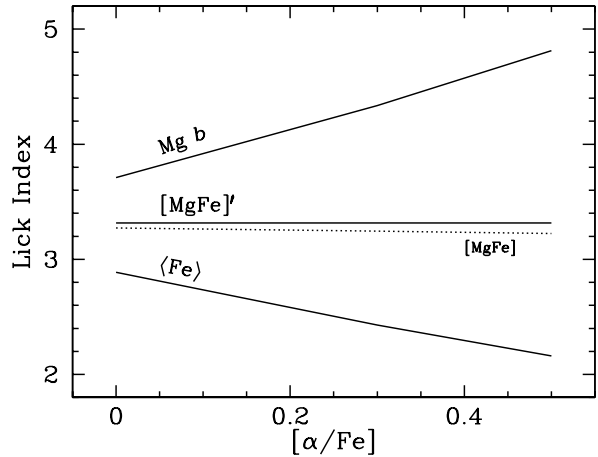,width=\linewidth}
\caption{Lick indices as function of \aFe\ ratio at fixed total
metallicity. Plotted are the indices \Mgb, $\Fe=\frac{1}{2}({\rm
Fe5270}+{\rm Fe5335})$, and $[{\rm MgFe}]^{\prime}\equiv\sqrt{\Mgb\
(0.72\cdot {\rm Fe5270}+0.28\cdot{\rm Fe5335})}$ of 12~Gyr old SSP
models with solar metallicity. The dotted line is the index [MgFe] as
defined by \citet{G93}.}
\label{fig:afemgfe}
\end{figure}

In Fig.~\ref{fig:afemgfe} we plot the Lick indices \Mgb\ and \Fe\ of
our SSP models as functions of \aFe\ ratio at fixed total metallicity.
Models with age 12~Gyr and solar metallicity are shown.  \Mgb\
increases and \Fe\ decreases with increasing \aFe.  The index [MgFe]
as defined by \citet{G93} is the dotted line. Although being only very
little sensitive to \aFe, still [MgFe] slightly decreases with
increasing \aFe. 

From Table~\ref{tab:indices} we know that Fe5270 responds less to
\aFe\ ratio changes than Fe5335. Decreasing the weight of Fe5335 in
the definition of [MgFe] thus helps to remove the sensitivity to \aFe.
We therefore slightly modify the definition of [MgFe] and define the
new index
\begin{equation}
{\rm [MgFe]}^{\prime}\equiv \sqrt{\Mgb\ (0.72\cdot {\rm
Fe5270}+0.28\cdot {\rm Fe5335})}\ .
\end{equation}
In Fig.~\ref{fig:afemgfe} it is shown that ${\rm [MgFe]}^{\prime}$ is
indeed completely independent of \aFe. We note that this behaviour is
almost independent of the adopted age and metallicity.  Therefore
[MgFe]$^\prime$ serves best as a tracer of the total metallicity of
stellar populations.


\section{Conclusions}
\label{sec:summary}
We present a comprehensive set of new generation stellar population
models of Lick absorption line indices, which for the first time
include abundance ratios different from solar. We computed the 21 Lick
indices \CNone, \CNtwo, Ca4227, G4300, Fe4383, Ca4455, Fe4531, \FeC,
\Hb, Fe5015, \Mgone, \Mgtwo, \Mgb, Fe5270, Fe5335, Fe5406, Fe5709,
Fe5782, Na~D, \TiOone, and \TiOtwo\ in the wavelength range $4000\la
\lambda\la 6500$~\AA. Models are provided with:
$[\aFe]=0.0,\:0.3,\:0.5$, $[\aCa]=-0.1,\:0.0,\:0.2,\:0.5$, and
$[\aN]=-0.5,\:0.0$; ages from 1 to 15~Gyr; total metallicities from
1/200 to 3.5 solar ($-2.25\leq [\ZH]\leq 0.67$).

The models are based on the evolutionary synthesis technique described
in \citet{Ma98}. The \aFe\ enhanced mixtures are obtained by
increasing the abundances of $\alpha$-group elements and by decreasing
the abundances of the Fe-peak elements, such that total metallicity is
conserved.  The impact from these element abundance variations on the
absorption line indices is taken from \citet{TB95}, using an extension
of the method introduced by \citet{Traetal00a}. Most importantly, we
take into account that the empirical stellar libraries used to compute
model indices follow the chemical enrichment history of the Milky Way,
and are therefore biased towards super-solar \aFe\ ratios at sub-solar
metallicities. We corrected for this bias, so that the models
presented here have well-defined \aFe\ ratios at all metallicities.

We take particular care at calibrating the models with galactic
globular clusters, for which ages, metallicities, and element
abundance ratios are known from independent sources. Our \aFe\
enhanced models with $[\aFe]=0.3$ (and 12~Gyr age) perfectly reproduce
the positions of the globular cluster data in the \Mgb-\Fe\ diagram up
to solar metallicities \citep[see also][]{Maretal02}.  The total
metallicities for the sample clusters that we derive from these
indices are in excellent agreement with the \citet{ZW84} metallicity
scale. We point out that the latter most likely reflects total
metallicity rather than iron abundance, because it is obtained
essentially by averaging the abundances derived from the Mg triplet
near 5175~\AA\ and the Fe blend at 5270~\AA\
\citep{Cohen83,ZW84}. This aspect needs to be emphasized, as with the
\aFe\ enhanced models we are now in the position to distinguish total
metallicity [\ZH] and iron abundance [Fe/H].

By means of our calibrated \aFe\ enhanced models, we confirm that the
index [MgFe], suggested by \citet{G93} to balance \aFe\ ratio effects,
is almost independent of \aFe. As it modestly decreases with
increasing \aFe, however, we define the slightly modified index
\[
{\rm [MgFe]}^{\prime}\equiv \sqrt{\Mgb\ (0.72\cdot {\rm
Fe5270}+0.28\cdot {\rm Fe5335})}\ , 
\]
which is completely independent of \aFe, and hence an even better
tracer of total metallicity. We further show that the linear
correlation between \Mgtwo\ and metallicity at old ages derived
empirically by \citet{BH90} is valid up to $\sim 1/3$ solar
metallicity, but underpredicts \Mgtwo\ indices at metallicities above
that threshold.

It turns out to be hard to find indices that correlate with \aFe\ as
well as the intensively studied indices \Mgone, \Mgtwo, and
\Mgb. Promising alternatives are the blue indices \CNone\ and \CNtwo\
that also increase with increasing \aFe\ ratio, mainly because of an
anti-correlation with Fe abundance. With the caveat that \CNone\ and
\CNtwo\ are additionally sensitive to C and N abundances, they can be
regarded to be complementary to the indices \Mgone, \Mgtwo, and \Mgb.
Alternatives to the iron indices Fe5270 and Fe5335, the strengths of
which decrease with increasing \aFe\ ratio, are easier to find. The
best cases are the indices Fe4383, Fe4531, Fe5015, and Fe5709.

The indices \CNone, \CNtwo, and Ca4227 of globular clusters are very
interesting, particular cases.  We find that the relatively strong CN
features observed in globular clusters require models in which
nitrogen is enhanced by a factor three relative to the
$\alpha$-elements, hence $[\aN]=-0.5$.  This is in agreement with
early suggestions by \citet{DGC83} and \citet{Renzini83}, that stars
in globular clusters may accrete carbon and/or nitrogen enriched
ejecta from the surrounding AGB stars \citep{RV81}.  The good
calibration of other indices like \Mgone, \Mgb\ or \Fe\ is not
affected by a variation of the \aN\ ratio, as these indices are not
sensitive to nitrogen abundance.  We note that an enhancement of
carbon abundance, instead, would lead to serious inconsistencies with
\Mgone.  Interestingly, also Ca4227 is sensitive to nitrogen
abundance, and the globular cluster data of this index are also best
reproduced by the model with increased nitrogen abundance.

\medskip
To conclude, the stellar population models presented here make it
possible, for the first time, to study in detail individual element
abundance ratios of unresolved stellar populations. In particular,
total metallicity is now a well-defined quantity. In an accompanying
paper (D.~Thomas et al., in preparation), we use these models to
derive quantitatively \aCa\ and \CaFe\ ratios of the stellar
populations in elliptical galaxies from their Ca4227, \Mgb, and \Fe\
indices.  Interesting for galaxy formation will also be to investigate
element abundance ratios of galaxies at earlier stages of their
evolution. On the basis of the calibration carried out in this paper,
we suggest that the combination of the blue Lick indices \CNone\ and
Fe4383 may be best suited to estimate \aFe\ ratios of objects at
redshifts $z\sim 1$.


\section*{Acknowledgments}
We would like to thank Laura Greggio for the numerous, very fruitful
discussions.  DT and CM thank Claudia Mendes de Oliveira, Beatriz
Barbuy, and the members of the Instituto Astronomico e Geofisico of
S\~ao Paulo for their kind hospitality. Santi Cassisi is acknowledged
for providing a large set of stellar evolutionary tracks. We thank B.\
Barbuy, A.\ Renzini, and S.\ Trager for very interesting and
stimulating discussions. We acknowledge the anonymous referee. The
"Sonderforschungsbereich 375-95 f\"ur Astro-Teilchenphysik" of the
Deutsche Forschungsgemeinschaft, the BMBF, the DAAD, and the FAPESP
are acknowledged for financial support.


\small

\normalsize

\bsp
\label{lastpage}


\appendix

\section{\boldmath Models with variable \aFe\ ratios}
\label{app:models}
In Tables~A1 to A3 we provide the SSP models for the 21 Lick indices
\CNone, \CNtwo, Ca4227, G4300, Fe4383, Ca4455, Fe4531, \FeC, \Hb,
Fe5015, \Mgone, \Mgtwo, \Mgb, Fe5270, Fe5335, Fe5406, Fe5709, Fe5782,
Na~D, \TiOone, and \TiOtwo. The models comprise the ages $1$ to
$15$~Gyr, the total metallicities $-2.25\leq [\ZH]\leq 0.67$, and the
\aFe\ ratios $[\aFe]=0.0$ (Table~A1), $[\aFe]=0.3$ (Table~A2), and
$[\aFe]=0.5$ (Table~A3). These models have $[\aN]=0$ and
$[\aCa]=0$. In a more comprehensive form (i.e., finer grid in ages and
\aFe\ ratios), they are also available electronically via ftp at {\tt
ftp.mpe.mpg.de} in the directory {\tt people/dthomas/SSPs} or via WWW
at {\tt ftp://ftp.mpe.mpg.de/people/dthomas/SSPs}.

\section{\boldmath N enhanced models}
\label{app:N-models}
Table~B1 contains SSP models of the indices \CNone, \CNtwo, and Ca4227
in which the element nitrogen is enhanced by a factor 3
($[\aN]=-0.5$). The models comprise the same ages, metallicities, and
\aFe\ ratios as in Tables~A1 to A3.

\section{\boldmath Models with variable \aCa\ ratios}
\label{app:Ca-models}
Table~C1 contains SSP models of the index Ca4227 with the \aCa\ ratios
$[\aCa]=-0.1,\:0.2,\:0.5$. Note that Ca4227 is very insensitive to
\aFe.

\clearpage
\setcounter{section}{1}
\begin{table*}
\caption{$[\aFe]=0.0$.}
\begin{tabular}{rrrrrrrrrrrr}
\hline\hline
\multicolumn{1}{c}{Age} & \multicolumn{1}{c}{[\ZH]} & \multicolumn{1}{c}{\CNone} & \multicolumn{1}{c}{\CNtwo} & \multicolumn{1}{c}{Ca4227} & \multicolumn{1}{c}{G4300} & \multicolumn{1}{c}{Fe4383} & \multicolumn{1}{c}{Ca4455} & \multicolumn{1}{c}{Fe4531} & \multicolumn{1}{c}{\FeC} & \multicolumn{1}{c}{\Hb}  & \multicolumn{1}{c}{Fe5015} \\
\multicolumn{1}{c}{(1)} & \multicolumn{1}{c}{(2)}   & \multicolumn{1}{c}{(3)}    & \multicolumn{1}{c}{(4)}    & \multicolumn{1}{c}{(5)}    & \multicolumn{1}{c}{(6)}   & \multicolumn{1}{c}{(7)}    & \multicolumn{1}{c}{(8)}    & \multicolumn{1}{c}{(9)}    & \multicolumn{1}{c}{(10)} & \multicolumn{1}{c}{(11)} & \multicolumn{1}{c}{(12)}   \\
\hline
$ 1$ & $-2.25$ & $-0.2632$ & $-0.1536$ & $-0.4260$ & $-4.3872$ & $-2.8764$ & $-0.1924$ & $-0.8226$ & $ 1.6082$ & $ 6.7915$ & $-0.0725$\\
$ 2$ & $-2.25$ & $-0.2033$ & $-0.1214$ & $-0.2038$ & $-2.4775$ & $-1.8406$ & $-0.0604$ & $-0.3240$ & $ 0.9274$ & $ 4.8015$ & $ 0.3634$\\
$ 3$ & $-2.25$ & $-0.1871$ & $-0.1137$ & $-0.1132$ & $-1.9934$ & $-1.4889$ & $-0.0243$ & $-0.1584$ & $ 0.6396$ & $ 4.3819$ & $ 0.4838$\\
$ 4$ & $-2.25$ & $-0.1740$ & $-0.1077$ & $-0.0440$ & $-1.6085$ & $-1.2146$ & $-0.0036$ & $-0.0539$ & $ 0.4111$ & $ 4.0630$ & $ 0.5427$\\
$ 5$ & $-2.25$ & $-0.1621$ & $-0.1021$ & $ 0.0173$ & $-1.2265$ & $-0.9724$ & $ 0.0149$ & $ 0.0312$ & $ 0.2273$ & $ 3.7640$ & $ 0.5811$\\
$ 6$ & $-2.25$ & $-0.1491$ & $-0.0941$ & $ 0.0661$ & $-0.8894$ & $-0.6061$ & $ 0.0511$ & $ 0.1253$ & $ 0.1278$ & $ 3.5048$ & $ 0.6476$\\
\hline
\end{tabular}
\end{table*}

\clearpage
\begin{table*}
\contcaption{$[\aFe]=0.0$.}
\begin{tabular}{rrrrrrrrrrrrr}
\hline\hline
\multicolumn{1}{c}{Age} & \multicolumn{1}{c}{[\ZH]} & \multicolumn{1}{c}{\Mgone} & \multicolumn{1}{c}{\Mgtwo} & \multicolumn{1}{c}{\Mgb} & \multicolumn{1}{c}{Fe5270} & \multicolumn{1}{c}{Fe5335} & \multicolumn{1}{c}{Fe5406} & \multicolumn{1}{c}{Fe5709} & \multicolumn{1}{c}{Fe5782} & \multicolumn{1}{c}{NaD} & \multicolumn{1}{c}{\TiOone} & \multicolumn{1}{c}{\TiOtwo}  \\
\multicolumn{1}{c}{(1)} & \multicolumn{1}{c}{(2)}   & \multicolumn{1}{c}{(3)}    & \multicolumn{1}{c}{(4)}    & \multicolumn{1}{c}{(5)}    & \multicolumn{1}{c}{(6)}   & \multicolumn{1}{c}{(7)}    & \multicolumn{1}{c}{(8)}    & \multicolumn{1}{c}{(9)}    & \multicolumn{1}{c}{(10)} & \multicolumn{1}{c}{(11)} & \multicolumn{1}{c}{(12)} & \multicolumn{1}{c}{(13)}   \\
\hline
$ 1$ & $-2.25$ & $ 0.0115$ & $-0.0025$ & $ 0.1794$ & $ 0.1715$ & $ 0.2673$ & $-0.1296$ & $-0.0573$ & $-0.4248$ & $ 0.8827$ & $ 0.0140$ & $-0.0048$\\
$ 2$ & $-2.25$ & $ 0.0023$ & $ 0.0094$ & $ 0.3366$ & $ 0.3408$ & $ 0.4326$ & $ 0.0056$ & $-0.0074$ & $-0.3342$ & $ 0.8046$ & $ 0.0132$ & $-0.0037$\\
$ 3$ & $-2.25$ & $ 0.0012$ & $ 0.0123$ & $ 0.3632$ & $ 0.4333$ & $ 0.4781$ & $ 0.0519$ & $ 0.0094$ & $-0.3070$ & $ 0.8133$ & $ 0.0125$ & $-0.0039$\\
$ 4$ & $-2.25$ & $ 0.0003$ & $ 0.0167$ & $ 0.4429$ & $ 0.4910$ & $ 0.5023$ & $ 0.0725$ & $ 0.0119$ & $-0.2974$ & $ 0.8275$ & $ 0.0121$ & $-0.0042$\\
$ 5$ & $-2.25$ & $ 0.0000$ & $ 0.0205$ & $ 0.5125$ & $ 0.5421$ & $ 0.5246$ & $ 0.0835$ & $ 0.0140$ & $-0.2900$ & $ 0.8497$ & $ 0.0117$ & $-0.0045$\\
$ 6$ & $-2.25$ & $-0.0011$ & $ 0.0230$ & $ 0.5364$ & $ 0.6168$ & $ 0.5902$ & $ 0.1264$ & $ 0.0248$ & $-0.2677$ & $ 0.8524$ & $ 0.0114$ & $-0.0047$\\
\hline
\end{tabular}
\end{table*}

\clearpage
\begin{table*}
\caption{$[\aFe]=0.3$.}
\begin{tabular}{rrrrrrrrrrrr}
\hline\hline
\multicolumn{1}{c}{Age} & \multicolumn{1}{c}{[\ZH]} & \multicolumn{1}{c}{\CNone} & \multicolumn{1}{c}{\CNtwo} & \multicolumn{1}{c}{Ca4227} & \multicolumn{1}{c}{G4300} & \multicolumn{1}{c}{Fe4383} & \multicolumn{1}{c}{Ca4455} & \multicolumn{1}{c}{Fe4531} & \multicolumn{1}{c}{\FeC} & \multicolumn{1}{c}{\Hb}  & \multicolumn{1}{c}{Fe5015} \\
\multicolumn{1}{c}{(1)} & \multicolumn{1}{c}{(2)}   & \multicolumn{1}{c}{(3)}    & \multicolumn{1}{c}{(4)}    & \multicolumn{1}{c}{(5)}    & \multicolumn{1}{c}{(6)}   & \multicolumn{1}{c}{(7)}    & \multicolumn{1}{c}{(8)}    & \multicolumn{1}{c}{(9)}    & \multicolumn{1}{c}{(10)} & \multicolumn{1}{c}{(11)} & \multicolumn{1}{c}{(12)}   \\
\hline
$ 1$ & $-2.25$ & $-0.2597$ & $-0.1509$ & $-0.4300$ & $-4.3330$ & $-3.0443$ & $-0.1991$ & $-0.8268$ & $ 1.5056$ & $ 6.9087$ & $-0.0714$\\
$ 2$ & $-2.25$ & $-0.2023$ & $-0.1210$ & $-0.2076$ & $-2.4231$ & $-1.9632$ & $-0.0649$ & $-0.3439$ & $ 0.8304$ & $ 4.9274$ & $ 0.3419$\\
$ 3$ & $-2.25$ & $-0.1864$ & $-0.1137$ & $-0.1142$ & $-1.9177$ & $-1.6542$ & $-0.0261$ & $-0.1779$ & $ 0.5513$ & $ 4.5009$ & $ 0.4577$\\
$ 4$ & $-2.25$ & $-0.1734$ & $-0.1078$ & $-0.0414$ & $-1.5309$ & $-1.3888$ & $-0.0044$ & $-0.0702$ & $ 0.3322$ & $ 4.1729$ & $ 0.5150$\\
$ 5$ & $-2.25$ & $-0.1614$ & $-0.1022$ & $ 0.0234$ & $-1.1476$ & $-1.1470$ & $ 0.0150$ & $ 0.0177$ & $ 0.1563$ & $ 3.8651$ & $ 0.5525$\\
$ 6$ & $-2.25$ & $-0.1483$ & $-0.0942$ & $ 0.0755$ & $-0.8130$ & $-0.8060$ & $ 0.0526$ & $ 0.1132$ & $ 0.0644$ & $ 3.5966$ & $ 0.6159$\\
\hline
\end{tabular}
\end{table*}

\clearpage
\begin{table*}
\contcaption{$[\aFe]=0.3$.}
\begin{tabular}{rrrrrrrrrrrrr}
\hline\hline
\multicolumn{1}{c}{Age} & \multicolumn{1}{c}{[\ZH]} & \multicolumn{1}{c}{\Mgone} & \multicolumn{1}{c}{\Mgtwo} & \multicolumn{1}{c}{\Mgb} & \multicolumn{1}{c}{Fe5270} & \multicolumn{1}{c}{Fe5335} & \multicolumn{1}{c}{Fe5406} & \multicolumn{1}{c}{Fe5709} & \multicolumn{1}{c}{Fe5782} & \multicolumn{1}{c}{NaD} & \multicolumn{1}{c}{\TiOone} & \multicolumn{1}{c}{\TiOtwo}  \\
\multicolumn{1}{c}{(1)} & \multicolumn{1}{c}{(2)}   & \multicolumn{1}{c}{(3)}    & \multicolumn{1}{c}{(4)}    & \multicolumn{1}{c}{(5)}    & \multicolumn{1}{c}{(6)}   & \multicolumn{1}{c}{(7)}    & \multicolumn{1}{c}{(8)}    & \multicolumn{1}{c}{(9)}    & \multicolumn{1}{c}{(10)} & \multicolumn{1}{c}{(11)} & \multicolumn{1}{c}{(12)} & \multicolumn{1}{c}{(13)}   \\
\hline
$ 1$ & $-2.25$ & $ 0.0141$ & $-0.0026$ & $ 0.2098$ & $ 0.1039$ & $ 0.2060$ & $-0.1542$ & $-0.0526$ & $-0.4449$ & $ 0.8640$ & $ 0.0137$ & $-0.0053$\\
$ 2$ & $-2.25$ & $ 0.0049$ & $ 0.0101$ & $ 0.3873$ & $ 0.2550$ & $ 0.3434$ & $-0.0306$ & $-0.0067$ & $-0.3484$ & $ 0.7862$ & $ 0.0129$ & $-0.0041$\\
$ 3$ & $-2.25$ & $ 0.0035$ & $ 0.0131$ & $ 0.4097$ & $ 0.3476$ & $ 0.3878$ & $ 0.0146$ & $ 0.0087$ & $-0.3190$ & $ 0.8009$ & $ 0.0122$ & $-0.0043$\\
$ 4$ & $-2.25$ & $ 0.0023$ & $ 0.0177$ & $ 0.4998$ & $ 0.4090$ & $ 0.4065$ & $ 0.0395$ & $ 0.0110$ & $-0.3081$ & $ 0.8205$ & $ 0.0118$ & $-0.0046$\\
$ 5$ & $-2.25$ & $ 0.0020$ & $ 0.0217$ & $ 0.5789$ & $ 0.4634$ & $ 0.4238$ & $ 0.0566$ & $ 0.0128$ & $-0.2998$ & $ 0.8473$ & $ 0.0114$ & $-0.0049$\\
$ 6$ & $-2.25$ & $ 0.0009$ & $ 0.0244$ & $ 0.6040$ & $ 0.5357$ & $ 0.4754$ & $ 0.0969$ & $ 0.0226$ & $-0.2749$ & $ 0.8532$ & $ 0.0111$ & $-0.0051$\\
\hline
\end{tabular}
\end{table*}

\clearpage
\begin{table*}
\caption{$[\aFe]=0.5$.}
\begin{tabular}{rrrrrrrrrrrr}
\hline\hline
\multicolumn{1}{c}{Age} & \multicolumn{1}{c}{[\ZH]} & \multicolumn{1}{c}{\CNone} & \multicolumn{1}{c}{\CNtwo} & \multicolumn{1}{c}{Ca4227} & \multicolumn{1}{c}{G4300} & \multicolumn{1}{c}{Fe4383} & \multicolumn{1}{c}{Ca4455} & \multicolumn{1}{c}{Fe4531} & \multicolumn{1}{c}{\FeC} & \multicolumn{1}{c}{\Hb}  & \multicolumn{1}{c}{Fe5015} \\
\multicolumn{1}{c}{(1)} & \multicolumn{1}{c}{(2)}   & \multicolumn{1}{c}{(3)}    & \multicolumn{1}{c}{(4)}    & \multicolumn{1}{c}{(5)}    & \multicolumn{1}{c}{(6)}   & \multicolumn{1}{c}{(7)}    & \multicolumn{1}{c}{(8)}    & \multicolumn{1}{c}{(9)}    & \multicolumn{1}{c}{(10)} & \multicolumn{1}{c}{(11)} & \multicolumn{1}{c}{(12)}   \\
\hline                                                                                                                                   

$ 1$ & $-2.25$ & $-0.2572$ & $-0.1490$ & $-0.4347$ & $-4.2937$ & $-3.1467$ & $-0.2036$ & $-0.8295$ & $ 1.4433$ & $ 6.9993$ & $-0.0706$\\
$ 2$ & $-2.25$ & $-0.2016$ & $-0.1207$ & $-0.2101$ & $-2.3830$ & $-2.0388$ & $-0.0679$ & $-0.3574$ & $ 0.7673$ & $ 5.0155$ & $ 0.3281$\\
$ 3$ & $-2.25$ & $-0.1860$ & $-0.1137$ & $-0.1152$ & $-1.8620$ & $-1.7522$ & $-0.0273$ & $-0.1911$ & $ 0.4939$ & $ 4.5837$ & $ 0.4408$\\
$ 4$ & $-2.25$ & $-0.1729$ & $-0.1079$ & $-0.0403$ & $-1.4737$ & $-1.4904$ & $-0.0049$ & $-0.0812$ & $ 0.2810$ & $ 4.2494$ & $ 0.4971$\\
$ 5$ & $-2.25$ & $-0.1608$ & $-0.1022$ & $ 0.0268$ & $-1.0895$ & $-1.2473$ & $ 0.0151$ & $ 0.0086$ & $ 0.1102$ & $ 3.9359$ & $ 0.5342$\\
$ 6$ & $-2.25$ & $-0.1477$ & $-0.0943$ & $ 0.0811$ & $-0.7567$ & $-0.9188$ & $ 0.0536$ & $ 0.1049$ & $ 0.0229$ & $ 3.6613$ & $ 0.5956$\\
\hline
\end{tabular}
\end{table*}

\clearpage
\begin{table*}
\contcaption{$[\aFe]=0.5$.}
\begin{tabular}{rrrrrrrrrrrrr}
\hline\hline
\multicolumn{1}{c}{Age} & \multicolumn{1}{c}{[\ZH]} & \multicolumn{1}{c}{\Mgone} & \multicolumn{1}{c}{\Mgtwo} & \multicolumn{1}{c}{\Mgb} & \multicolumn{1}{c}{Fe5270} & \multicolumn{1}{c}{Fe5335} & \multicolumn{1}{c}{Fe5406} & \multicolumn{1}{c}{Fe5709} & \multicolumn{1}{c}{Fe5782} & \multicolumn{1}{c}{NaD} & \multicolumn{1}{c}{\TiOone} & \multicolumn{1}{c}{\TiOtwo}  \\
\multicolumn{1}{c}{(1)} & \multicolumn{1}{c}{(2)}   & \multicolumn{1}{c}{(3)}    & \multicolumn{1}{c}{(4)}    & \multicolumn{1}{c}{(5)}    & \multicolumn{1}{c}{(6)}   & \multicolumn{1}{c}{(7)}    & \multicolumn{1}{c}{(8)}    & \multicolumn{1}{c}{(9)}    & \multicolumn{1}{c}{(10)} & \multicolumn{1}{c}{(11)} & \multicolumn{1}{c}{(12)} & \multicolumn{1}{c}{(13)}   \\
\hline
$ 1$ & $-2.25$ & $ 0.0162$ & $-0.0026$ & $ 0.2327$ & $ 0.0603$ & $ 0.1690$ & $-0.1685$ & $-0.0496$ & $-0.4578$ & $ 0.8483$ & $ 0.0136$ & $-0.0056$\\
$ 2$ & $-2.25$ & $ 0.0071$ & $ 0.0105$ & $ 0.4252$ & $ 0.2007$ & $ 0.2915$ & $-0.0511$ & $-0.0063$ & $-0.3577$ & $ 0.7711$ & $ 0.0127$ & $-0.0044$\\
$ 3$ & $-2.25$ & $ 0.0052$ & $ 0.0137$ & $ 0.4437$ & $ 0.2940$ & $ 0.3362$ & $-0.0062$ & $ 0.0082$ & $-0.3268$ & $ 0.7897$ & $ 0.0120$ & $-0.0046$\\
$ 4$ & $-2.25$ & $ 0.0039$ & $ 0.0184$ & $ 0.5414$ & $ 0.3583$ & $ 0.3519$ & $ 0.0215$ & $ 0.0104$ & $-0.3150$ & $ 0.8128$ & $ 0.0115$ & $-0.0049$\\
$ 5$ & $-2.25$ & $ 0.0034$ & $ 0.0226$ & $ 0.6275$ & $ 0.4151$ & $ 0.3664$ & $ 0.0422$ & $ 0.0121$ & $-0.3061$ & $ 0.8425$ & $ 0.0111$ & $-0.0052$\\
$ 6$ & $-2.25$ & $ 0.0024$ & $ 0.0253$ & $ 0.6532$ & $ 0.4865$ & $ 0.4103$ & $ 0.0809$ & $ 0.0211$ & $-0.2796$ & $ 0.8506$ & $ 0.0109$ & $-0.0053$\\
\hline
\end{tabular}
\end{table*}

\setcounter{section}{2}
\setcounter{table}{0}
\begin{table*}
\caption{Models with $[\aN]=-0.5$.}
\begin{tabular}{rrrrrrrrrrr}
\hline\hline
 &  & \multicolumn{3}{c}{$[\aFe]=0.0$} & \multicolumn{3}{c}{$[\aFe]=0.3$} & \multicolumn{3}{c}{$[\aFe]=0.5$} \\
\multicolumn{1}{c}{Age} & \multicolumn{1}{c}{[\ZH]} & \multicolumn{1}{c}{\CNone} & \multicolumn{1}{c}{\CNtwo} & \multicolumn{1}{c}{Ca4227} & \multicolumn{1}{c}{\CNone} & \multicolumn{1}{c}{\CNtwo} & \multicolumn{1}{c}{Ca4227} & \multicolumn{1}{c}{\CNone} & \multicolumn{1}{c}{\CNtwo} & \multicolumn{1}{c}{Ca4227}  \\
\multicolumn{1}{c}{(1)} & \multicolumn{1}{c}{(2)}   & \multicolumn{1}{c}{(3)}    & \multicolumn{1}{c}{(4)}    & \multicolumn{1}{c}{(5)}    & \multicolumn{1}{c}{(6)}   & \multicolumn{1}{c}{(7)}    & \multicolumn{1}{c}{(8)}    & \multicolumn{1}{c}{(9)}    & \multicolumn{1}{c}{(10)} & \multicolumn{1}{c}{(11)} \\
\hline
$ 1$ & $-2.25$ & $-0.2328$ & $-0.1274$ & $-0.4491$ & $-0.2298$ & $-0.1249$ & $-0.4521$ & $-0.2273$ & $-0.1226$ & $-0.4561$\\
$ 2$ & $-2.25$ & $-0.1882$ & $-0.1102$ & $-0.2044$ & $-0.1873$ & $-0.1099$ & $-0.2081$ & $-0.1863$ & $-0.1094$ & $-0.2106$\\
$ 3$ & $-2.25$ & $-0.1752$ & $-0.1060$ & $-0.1195$ & $-0.1745$ & $-0.1061$ & $-0.1201$ & $-0.1738$ & $-0.1060$ & $-0.1211$\\
$ 4$ & $-2.25$ & $-0.1629$ & $-0.1012$ & $-0.0509$ & $-0.1622$ & $-0.1014$ & $-0.0482$ & $-0.1615$ & $-0.1014$ & $-0.0470$\\
$ 5$ & $-2.25$ & $-0.1511$ & $-0.0961$ & $ 0.0100$ & $-0.1503$ & $-0.0963$ & $ 0.0161$ & $-0.1495$ & $-0.0962$ & $ 0.0195$\\
\hline
\end{tabular}
\label{tab:N-models}
\end{table*}

\setcounter{section}{3}
\setcounter{table}{0}
\begin{table*}
\caption{Models of Ca4227 with $[\aFe]=0.0$ and the variable \aCa\
ratios $[\aCa]=-0.1,\:0.2,\:0.5$.}
\begin{tabular}{rrrrr}
\hline\hline
\multicolumn{1}{c}{Age} & \multicolumn{1}{c}{[\ZH]} & \multicolumn{1}{c}{$-0.1$} & \multicolumn{1}{c}{$0.2$} & \multicolumn{1}{c}{$0.5$}\\
\hline
$ 1$ & $-2.25$ & $-0.3963$ & $-0.4728$ & $-0.5203$ \\
$ 2$ & $-2.25$ & $-0.2030$ & $-0.2088$ & $-0.2205$ \\
$ 3$ & $-2.25$ & $-0.1035$ & $-0.1301$ & $-0.1500$ \\
$ 4$ & $-2.25$ & $-0.0319$ & $-0.0635$ & $-0.0846$ \\
$ 5$ & $-2.25$ & $ 0.0312$ & $-0.0043$ & $-0.0261$ \\
$ 6$ & $-2.25$ & $ 0.0798$ & $ 0.0450$ & $ 0.0242$ \\
$ 8$ & $-2.25$ & $ 0.1647$ & $ 0.1149$ & $ 0.0813$ \\
\hline
\end{tabular}
\label{tab:Ca-models}
\end{table*}

\end{document}